\documentclass[pre,twocolumn,amssymb]{revtex4}
\usepackage{epsfig,amssymb,amsmath,ulem,color}
\pdfoutput=1
\begin{document}
\title{On the study of force-balance percolation}
\author{M. Jeng and J. M. Schwarz}
\affiliation{Department of Physics, Syracuse University, Syracuse, NY 13244}

\date{\today}

\begin{abstract}
We study models of correlated percolation where there are 
constraints on the occupation of sites that mimic force-balance, 
i.e. for a site to be stable requires occupied neighboring sites in all
four compass directions in two dimensions.
We prove
rigorously that $p_c<1$ for the two-dimensional models
studied. Numerical data indicate that the force-balance percolation
transition is discontinuous with a growing crossover length,
with perhaps the same form as the jamming percolation
models,
suggesting 
 the same underlying mechanism driving the transition in both cases. In other words, force-balance percolation and jamming percolation may indeed belong to the same universality class.
We
find a lower bound for the correlation length in the
connected phase and that the correlation function does not
appear to be a power law at the transition. Finally, we
study the dynamics of the culling procedure invoked to
obtain the force-balance configurations and find a dynamical
exponent similar to that found in sandpile models.

\end{abstract}

\maketitle

\section{Introduction}
\label{sec:intro}

Uncorrelated percolation and its associated geometric phase 
transition is arguably the most studied paradigm for a phase transition 
in a disordered system. Not only have physicists been able 
to nail down the universality class of percolation, more 
recently, mathematicians have been able to {\it rigorously} 
verify the universality class for at least one particular 
percolation model~\cite{SLE.percolation.1,SLE.percolation.2}. 
Even more recently, this
work has been extended to a second two-dimensional percolation 
model~\cite{SLE.percolation.extension}. Other two-dimensional 
models are expected to follow.

While uncorrelated percolation exhibits a continuous phase 
transition, there exists a new class of correlated percolation 
models that provably exhibit a {\it discontinuous} phase transition---a 
notable departure from the transition in uncorrelated
models~\cite{Knight,Knight.longer,Sandwich,TBF.response,Spiral.longer}. 
What do we mean by correlated percolation? We mean that 
there are constraints imposed on the occupation of sites 
such that correlations in the occupation arise regardless
of whether or not there is a transition. 

One of the simplest class of models of correlated percolation 
is $k$-core/bootstrap
percolation~\cite{kcore1,kcore2,SLC,network.guys,kcore.review}. It is defined as follows. 
Consider a regular lattice of coordination number $Z_{\rm 
max}$, and some integer $k$ with $2\le k< Z_{\rm max}$. Initially, 
sites are independently occupied with probability $p$. Then, all occupied sites with fewer than $k$
neighboring occupied
sites are eliminated. This decimation process is 
repeatedly applied to the surviving occupied sites, until 
all surviving sites (if any) have at least $k$ surviving 
neighbors. The surviving sites are called 
the $k$-core, and phases of the model are determined by 
the presence or absence of an infinite cluster of these 
survivors. 
The $k$-core percolation model has a number 
of physical realizations~\cite{kcore.apps}, including nonmagnetic 
impurities in a magnetic system~\cite{kcore1} and the glass transition via a kinetically-constrained spin-flip model known as the Fredrickson-Andersen 
model~\cite{fred1,fred2}.

Another well-known model of correlated percolation is the Kob-Andersen model~\cite{Kob.Andersen}, 
a particle-conserving counterpart to the Fredrickson-Andersen 
model. In the Kob-Andersen model, a particle can hop if 
and only if there are at least $m$ empty neighbors before 
and after a particle hop. 
As the density of particles increases, 
it becomes more difficult to hop, and the density of frozen 
particles increases, resulting in slower dynamics.
Eventually, the frozen particles percolate throughout the
system, resulting in a glass transition. 
Based 
on numerical analysis of the percolation of frozen particles, it was initially thought that the 
Kob-Andersen model exhibited a glass transition at a 
value of $p_c\approx 0.84$ in two dimensions. However, recent 
work by Toninelli, Biroli, and Fisher rigorously demonstrates 
that the thermodynamic $p_c$ is actually unity for the range 
of $m$ relevant to the glass transition~\cite{TBF.KA1,TBF.KA2}.

\begin{figure}[bth]
\begin{center}
\includegraphics{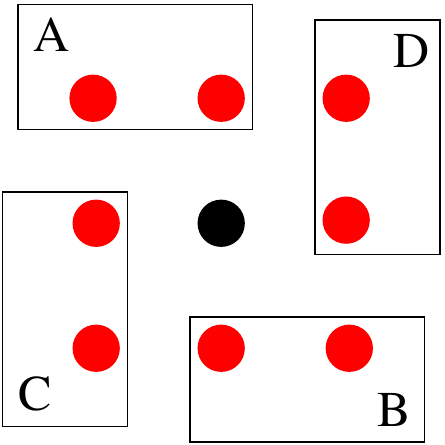}
\caption{Sets for the spiral model.}
\label{fig:spiral}
\end{center}
\end{figure}

Jamming percolation, a new class of two-dimensional 
correlated percolation models inspired by kinetically constrained 
models, was introduced by Toninelli, 
Biroli, and Fisher~\cite{Knight,Knight.longer,Spiral.longer}. These models consist 
of the following occupation constraints: a site can remain 
occupied only if there exists at least one occupied 
site in set A and one occupied site in set B, or one occupied 
site in set C and one occupied site in set D. All sets are 
disjoint from each other and contain two sites. See Fig.~\ref{fig:spiral} for
 the sets in a jamming percolation model 
called the spiral model. 
Jamming percolation models 
exhibit discontinuous phase transitions with a crossover 
length scale that diverges faster than any power law. It has been recently demonstrated that this 
behaviour prevails when more than four disjoint sets are 
introduced~\cite{JammingPerc.JSP}. While there are rigorous results for 
some models in this class, it is unknown how generic this 
transition is and whether or not there are other unusual, or
atypical, phase transitions of correlated percolation.

Here, we present a class of correlated percolation models denoted as 
force-balance percolation models. Force-balance percolation 
was originally introduced in Ref.~\cite{SLC} as a toy model 
for the jamming transition in finite dimensions. The jamming transition is a 
transition from a liquid-like to an amorphous solid-like state as some 
particular control parameter, such as the packing density, is varied. Examples of potentially related jamming systems are glass-forming 
liquids, colloidal suspensions, foams, emulsions, and granular matter~\cite{Liu.Nagel.PointJ}. Despite 
decades of study of the glass-forming liquids in particular, however, it is even unclear whether these transitions are true 
thermodynamic transitions or merely examples of kinetic arrest~\cite{glass.review}.

There has been some recent activity focusing on 
a zero-temperature jamming transition in a system of repulsive soft spheres as the packing density is increased. Numerical 
simulations by O'Hern, {\it et al.} considered repulsive soft 
particles in two and three dimensions~\cite{OHern.simulations,Langer}. 
For small packing fraction $\phi$, the particles easily arrange themselves 
so as not to overlap with any other particle, and the 
total potential energy thus vanishes. As $\phi$ is increased, 
there is a particular value of $\phi_c$ (Point J)
above which the particles can no longer avoid each other 
and the total potential energy becomes nonzero. The
system jams in that it develops nonzero static bulk and 
shear moduli above $\phi_c$. The average coordination number 
(the average number of overlapping neighbors per particle) 
jumps from $Z =0$ to $Z=Z_c$ at Point J,
and then rises with increasing packing fraction $\phi$ as 
$Z-Z_c \sim (\phi-\phi_c)^\beta$, where $\beta=0.5$. This behaviour was recently observed in a two-dimensional systems of glass beads~\cite{glass.beads.expt}. Furthermore, 
as $\phi$ approaches $\phi_c$ from above,
the singular part of the shear modulus vanishes with the 
exponent $\gamma=0.5$; more recent simulations also
find that there is a length scale that diverges with an 
exponent $\nu=0.25$~\cite{PointJ.more.sims}. 
So the transition at Point 
J appears to have characteristics of both first-order and 
second-order phase transitions: at the transition, $Z$ is 
discontinuous, but there
are nontrivial power laws. 

To understand this somewhat unusual transition, an 
analogy to $k$-core/bootstrap percolation was made in Ref.~\cite{SLC}. The scalar aspect of the principle of local mechanical stability, where particles need $d+1$ contacts, maps to a requirement of $k$ occupied neighbors. Surprising
agreement was found between  
the mean-field $k$-core percolation 
exponents and 
the repulsive soft sphere simulations. We note that the 
two- and three- dimensional simulations observed the same exponents, 
suggesting a possible critical dimension of two since logarithmic corrections would be difficult to determine~\cite{wyart}.

While there is agreement between the mean field $k$-core exponents and the 
low-dimensional repulsive soft sphere exponents, 
$k$-core in finite-dimensional spaces does not appear
to have such agreement.
Such systems seem to fall into one of two classes:
the transition is either continuous and in the same universality 
class as normal percolation, or it does not occur until 
$p_c=1$. In the first class, systems that allow finite clusters 
all exhibit continuous transitions~\cite{chavez,medeiros}. 
In the second class, large voids are very 
likely to grow, and in the infinite system limit, with probability 
one there will be at least one void that will grow to empty 
the entire system~\cite{kcore.odd.old.1,kcore.rigorous.1}. This prevents $k$-core percolation for 
any $p<1$. Unfortunately, neither category of $k$-core percolation
describes the nontrivial discontinuous transition observed 
in the finite-dimensional simulations and experiments of jamming.

In an effort to try to capture the behaviour of jamming in finite dimensions, force-balance percolation 
was introduced in Ref.~\cite{SLC}. In this model, the 
$k$-core constraint is retained, but the vectorial constraints of the 
principle of local mechanical stability are {\it mimicked} 
by creating culling rules that take into account where the neighboring 
particles are located. Loosely speaking, if there is a neighboring contact to 
one side of the particle, there must be at least one neighboring contact 
on the other side of the particle to allow for force balance. 

We must emphasize the force-balance percolation is a model with no explicit 
forces. We look only at connectivity, in contrast to models such as rigidity
percolation where repulsive and attractive forces are defined on, say,
a lattice of springs~\cite{rp1,rp2,rigidity.percolation.3,
rigidity.percolation.4}. However, the nature and possible
universality of the rigidity percolation transition is still up for
debate, despite decades of study. Since force-balance percolation is a
much simpler model there is ultimately a better chance of analyzing it
beyond numerics.

Here, we explore several versions of
force-balance percolation, both rigorously and numerically, to begin to 
answer the following questions:\\
\\
(0) Is there a force-balance percolation transition?\\
(1) If there is indeed a transition, 
what are its properties? Is it continuous?\\
(2) How generic is the transition among the various force-balance models?\\ 
(3) Is there a link between force-balance percolation and jamming percolation? Are they in the same universality class?\\

The paper is organized as follows. In section~\ref{sec:models} 
we review the force-balance percolation model introduced 
in Ref.~\cite{SLC}, and introduce two related models. We 
present in section~\ref{sec:rigorous} a rigorous proof that
the thermodynamic $p_c$ is less than unity for at least 
two of these models. Earlier work on $k$-core percolation 
misinterpreted numerical results, finding transitions in 
novel universality classes, with 
$p_c<1$~\cite{kcore.odd.old.1,kcore.odd.old.2,kcore.odd.old.3}, 
when in fact the models studied had 
$p_c=1$~\cite{kcore.rigorous.1,kcore.rigorous.2,kcore.rigorous.3}; 
our proof renders our subsequent interpretation of the numerical 
data presented in section~\ref{sec:numerical} on sounder footing. 
Finally, we close in section~\ref{sec:conclusions} with 
a summary of our findings and discuss their implications.


\section{Models}
\label{sec:models}

\begin{figure}[htb]
\begin{center}
\includegraphics[width=0.5\columnwidth]{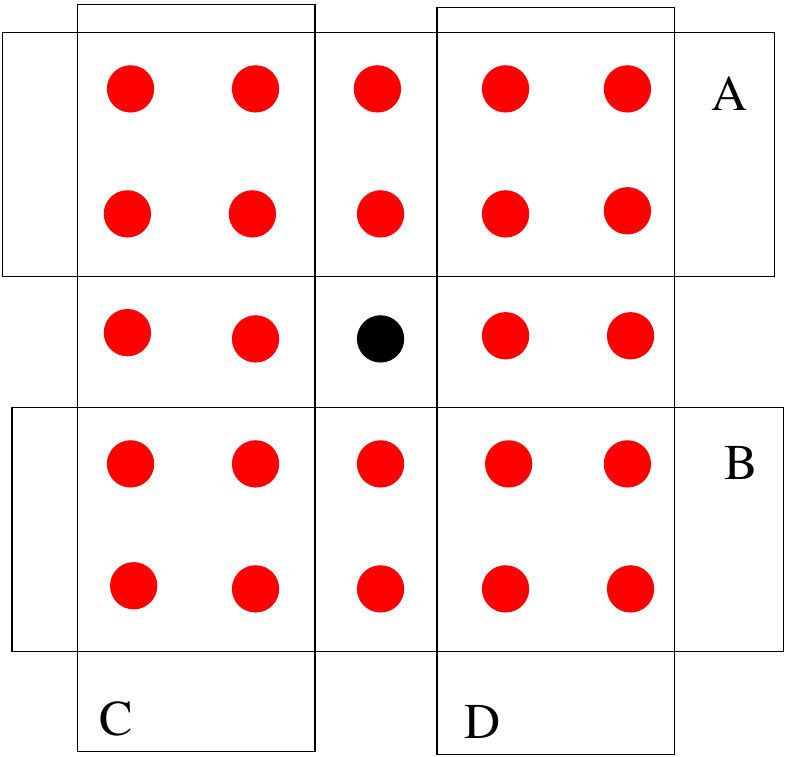}
\caption{Force-balance model on the 2$d$ square lattice 
with 24 nearest neighbors.}
\label{fig:z24.definition}
\end{center}
\end{figure}

\begin{figure}[htb]
\begin{center}
\includegraphics[width=0.4\columnwidth]{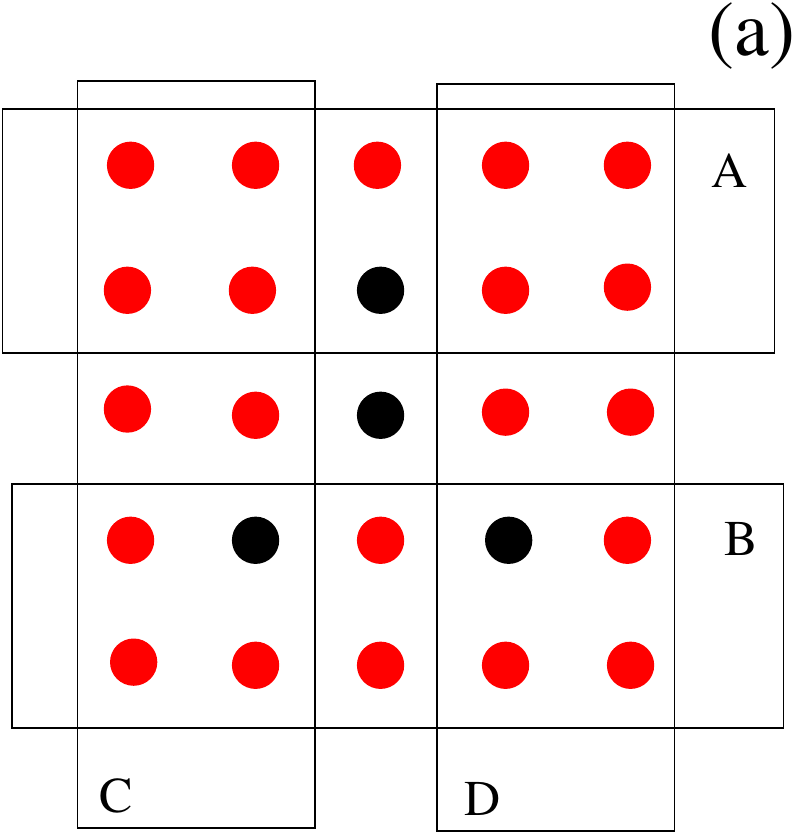}
\includegraphics[width=0.4\columnwidth]{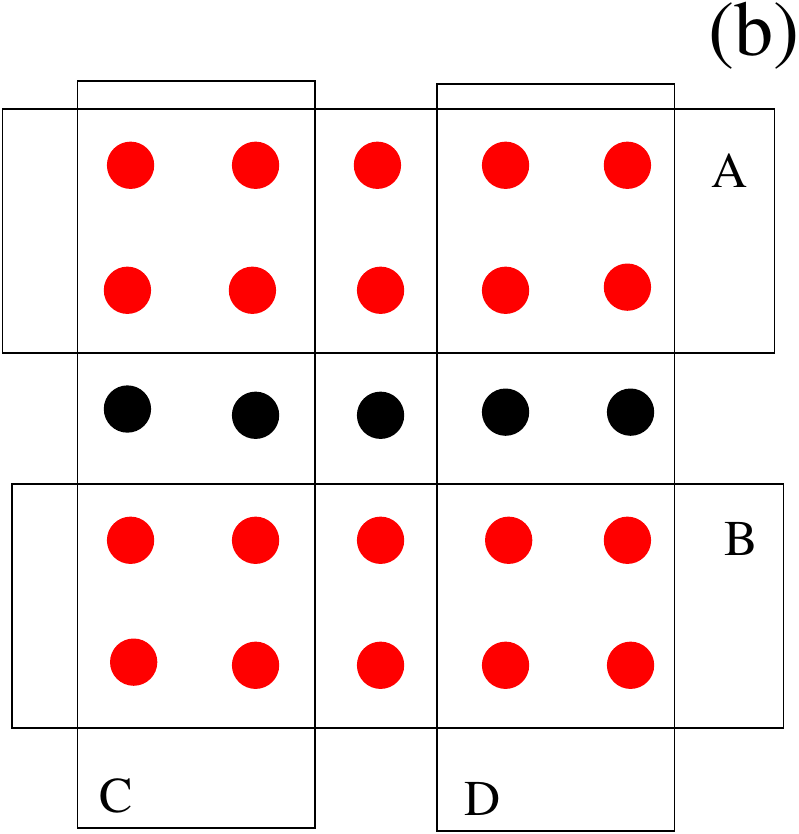}
\caption{Black sites denote occupied sites, and red sites 
denote unoccupied sites. (a) Allowed configuration. (b) 
Forbidden configuration. While the number of occupied nearest
neighbors is greater than three, the force-balance condition 
is violated.}
\label{fig:z24.allow}
\end{center}
\end{figure}

For the first force-balance percolation model, we begin 
with a two-dimensional square lattice. Each site neighbors 
all sites except itself within a 5x5 square---each site
therefore has 24 nearest neighbors. (For a 3x3 square, with the following rules, $p_c=1$). Since we are in two 
dimensions, we impose a 3-core constraint. However, we also 
impose the force-balance constraint, which is the following: 
there must be at least one occupied neighbor in set A, which 
in turn calls for at least one occupied neighbor in set 
B, and there must be at least one occupied neighbor in set 
C, which in turn calls for at least one occupied neighbor 
in set D. The four sets A, B, C, and D, are defined in 
Fig.~\ref{fig:z24.definition}. The force-balance constraint 
can be succinctly stated as: (A and B) and (C and D), where 
each letter X is short for ``at least one occupied site in 
set X''. Note that the force-balance constraint is defined 
in such a way such that vertical and/or horizontal lines of 
occupied particles are, by themselves, not stable. 
Fig.~\ref{fig:z24.allow}a 
demonstrates an allowed configuration and Fig.~\ref{fig:z24.allow}b 
demonstrates a prohibited configuration. 

To enforce the force-balance and $k$-core constraints, 
we initially occupy sites on the lattice with independent
occupation probabilities $p$, and then repeatedly 
remove occupied sites that violate either the $k$-core or 
force-balance constraints, until all remaining occupied 
sites are stable. Note that $p$ is the occupation density before 
culling, and generically differs from the final occupation 
density. 

The model is abelian. In other words, the 
final configuration after the culling process is independent 
of the order in which sites are culled. It can be done in parallel, 
or in series, or some combination thereof. One can define another force-balance-like model that allows for horizontal and vertical lines of occupied neighbors. However, 
such a model would be nonabelian. 
Nonabliean models are less desirable, because in such 
models the final results depend on the order in which sites
are culled. In our work, the culling procedure is merely an algorithm to achieve 
the force-balance percolation configuration. 

\begin{figure}[htb]
\begin{center}
\includegraphics[width=0.5\columnwidth]{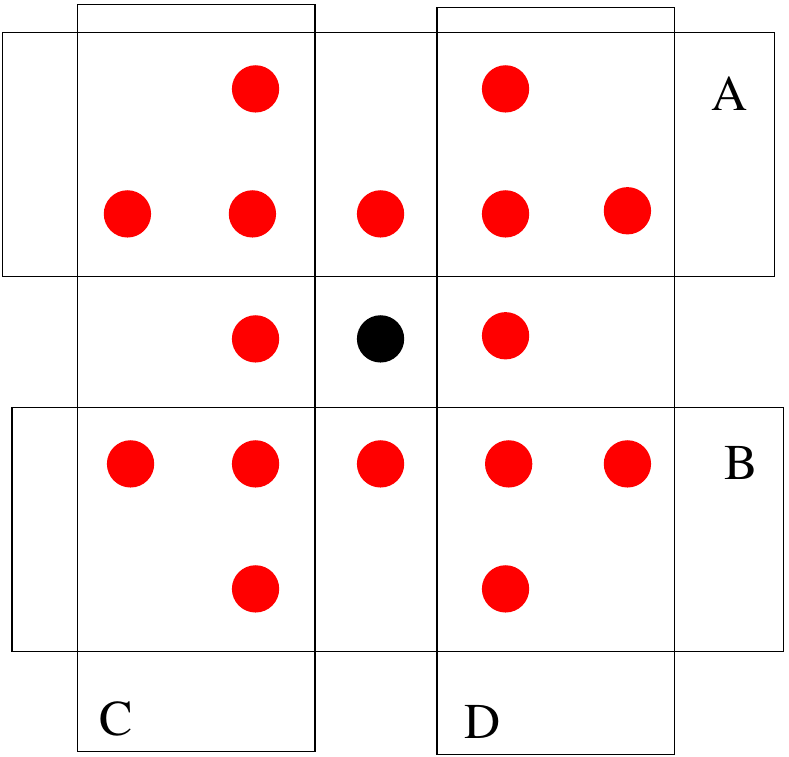}
\caption{A second force-balance model on the $2d$ square lattice 
with 16 nearest neighbors.}
\label{fig:z16.definition}
\end{center}
\end{figure}

To determine whether or not the behaviour observed in the 
original force-balance model is generic,
we define two additional abelian models. Our second model 
is defined on the square lattice, with 16 nearest neighbors, 
and quadrants as defined in Fig.~\ref{fig:z16.definition}. 
Again, $k=3$ and the force-balance constraint is the same 
as above: (A and B) and (C and D). This model is obviously 
similar to the first one, though the ratio of $k$ to $Z_{max}$ is 
different.

\begin{figure}[htb]
\begin{center}
\includegraphics[width=0.5\columnwidth]{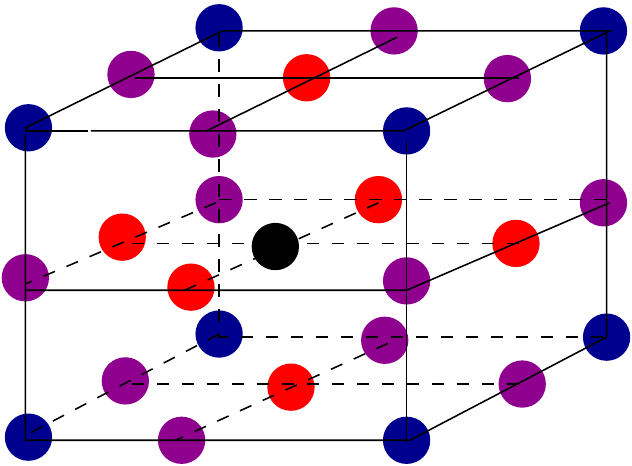}
\caption{Force-balance model on the $3d$ cubic lattice with 
26 nearest neighbors. A red site participates 
in only one of the six sets, a purple site 
in two of the six sets, and a blue site in three of the 
six sets.}
\label{fig:3d.fb.definition}
\end{center}
\end{figure}

The third model is a three-dimensional model with 26 nearest 
neighbors and six regions, color-coded in
Fig.~\ref{fig:3d.fb.definition}. Each of the six regions 
consists of a $3\times 3$ square on a face of the $3\times 
3\times 3$ cube centered on the site
whose stability is being analyzed.
For this model, we impose a 4-core constraint (since $d=3$), 
and the force-balance condition requires an ocupied site in 
each of the six regions. 

For comparison, along with numerical simulations of these 
three models, we also performed simulations on the spiral 
model. This model was introduced by Toninelli, {\it et al.}~\cite{TBF.response,Spiral.longer}, 
and they have proved that it undergoes a jamming transition,
and obtained rigorous results about the properties of the
transition~\cite{Spiral.longer}.

\section{Rigorous Results}
\label{sec:rigorous}

\subsection{Proof that \protect\boldmath$p_c<1$ in the 
\protect\boldmath$2d$ force-balance models}

\begin{figure}[htb]
\begin{center}
\includegraphics[width=0.5\columnwidth]{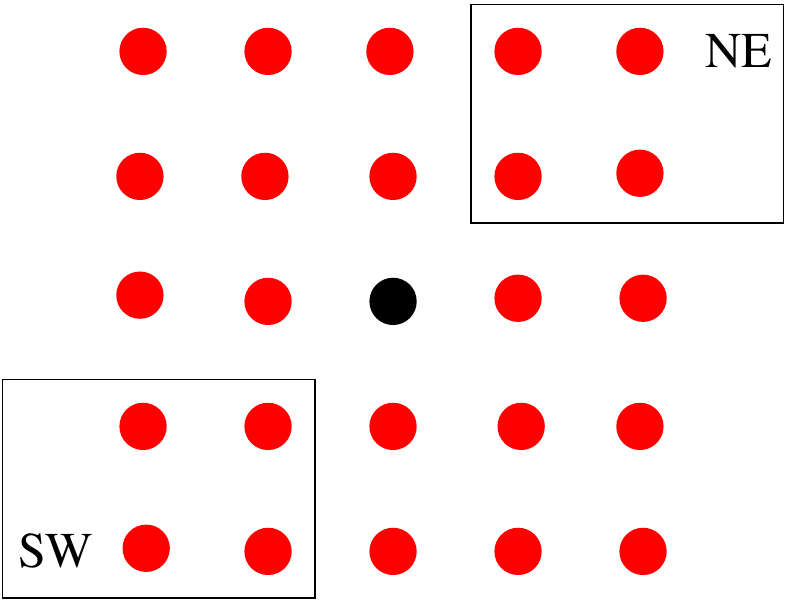}
\caption{$k=6$ NE-SW model.}
\label{fig:k6.NE.SW}
\end{center}
\end{figure}

To prove that $p_c<1$ for the initial force-balance model 
on the square lattice, we will demonstrate that $p_c<1$ 
for a more heavily constrained model with the same 24 nearest 
neighbors. In particular, we require (1) $k=6$, (2) at least 
one occupied neighbor in the 4-site region to the northeast,
and (3) at least one occupied neighbor in the 4-site region
to the southwest. See Fig.~\ref{fig:k6.NE.SW} 
for the two regions. Since
any sites stable under these conditions are automatically
stable in the force-balance model,
if $p_c<1$ for the $k=6$ NE-SW model,
then also $p_c<1$ for the force-balance model.

\begin{figure}[htb]
\begin{center}
\includegraphics[width=0.5\columnwidth]{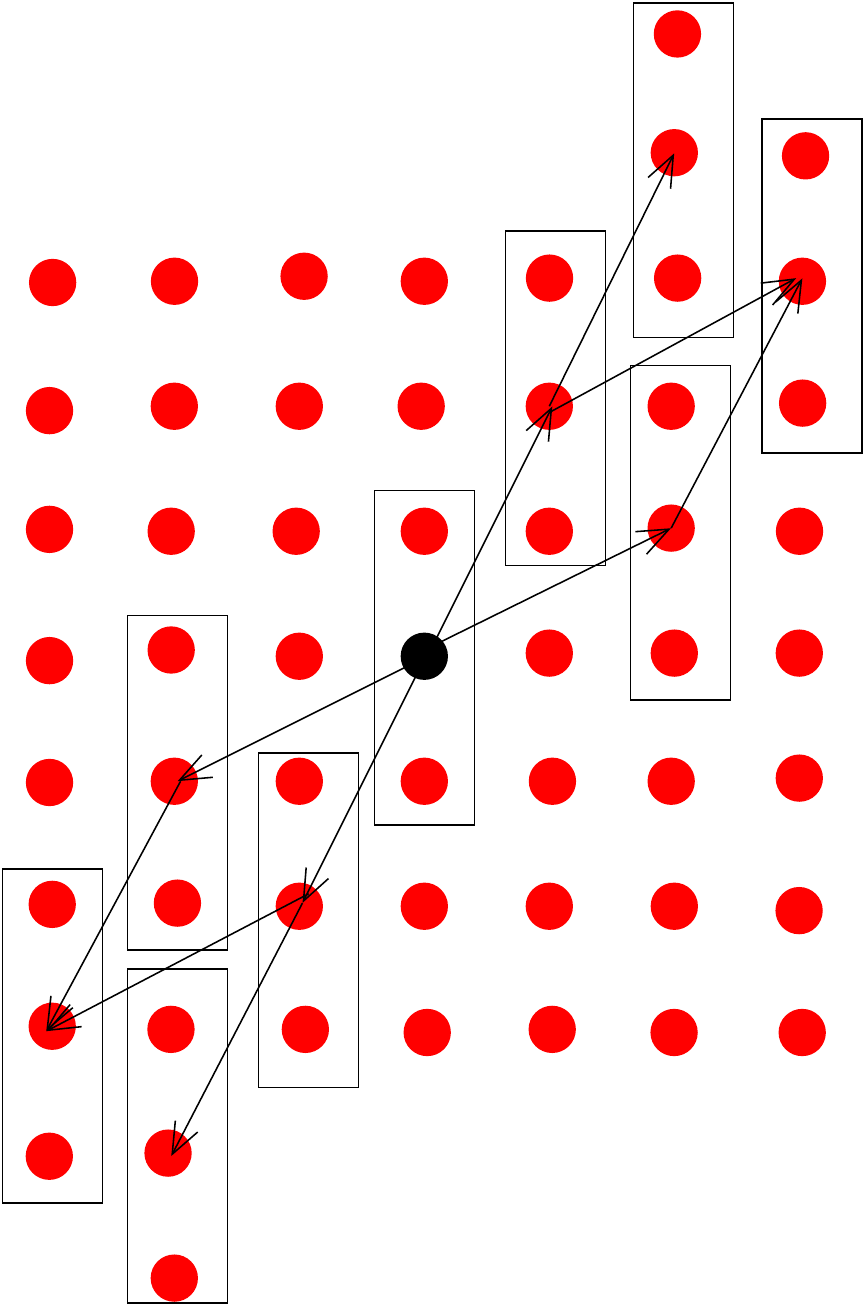}
\caption{Adjacent clusters of 3 sites in the NE-SW model.}
\label{fig:3clusters}
\end{center}
\end{figure}

\begin{figure}[htb]
\begin{center}
\includegraphics[width=0.2\columnwidth]{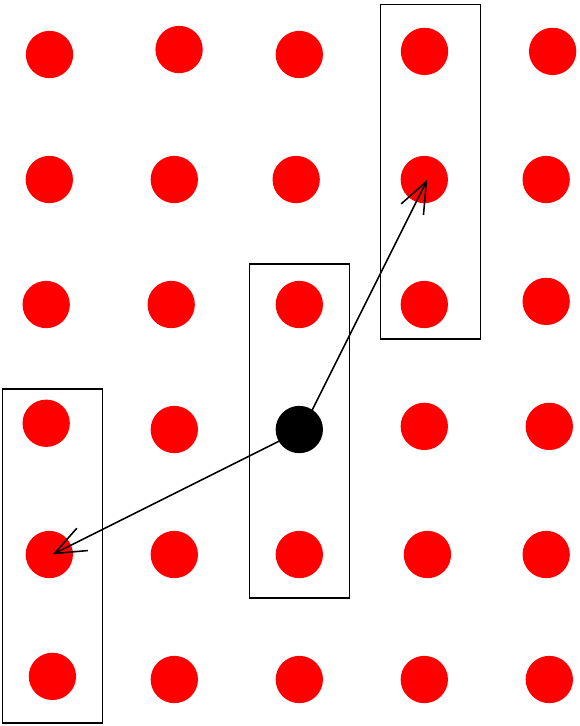}
\includegraphics[width=0.2\columnwidth]{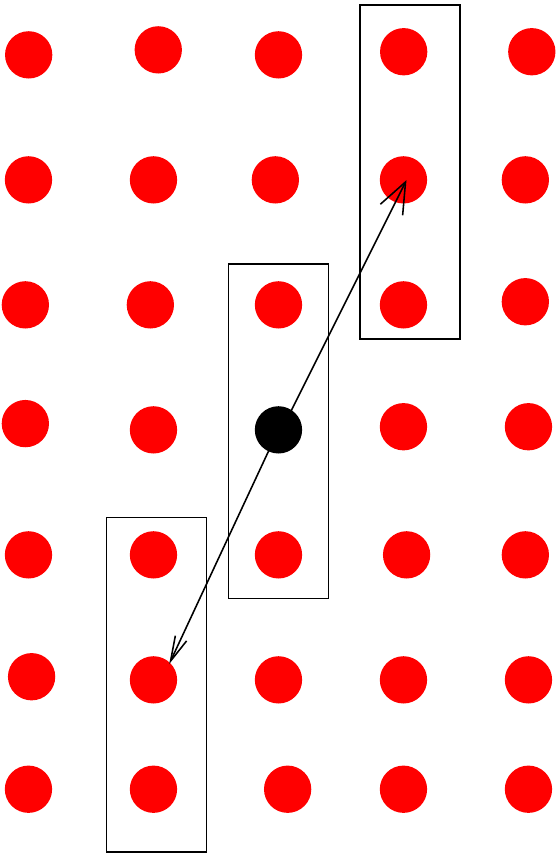}
\includegraphics[width=0.2\columnwidth]{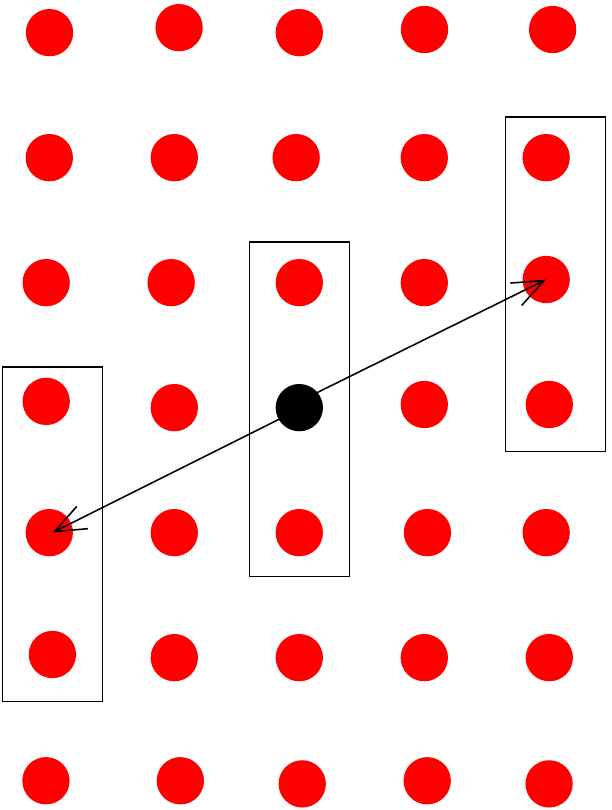}
\includegraphics[width=0.2\columnwidth]{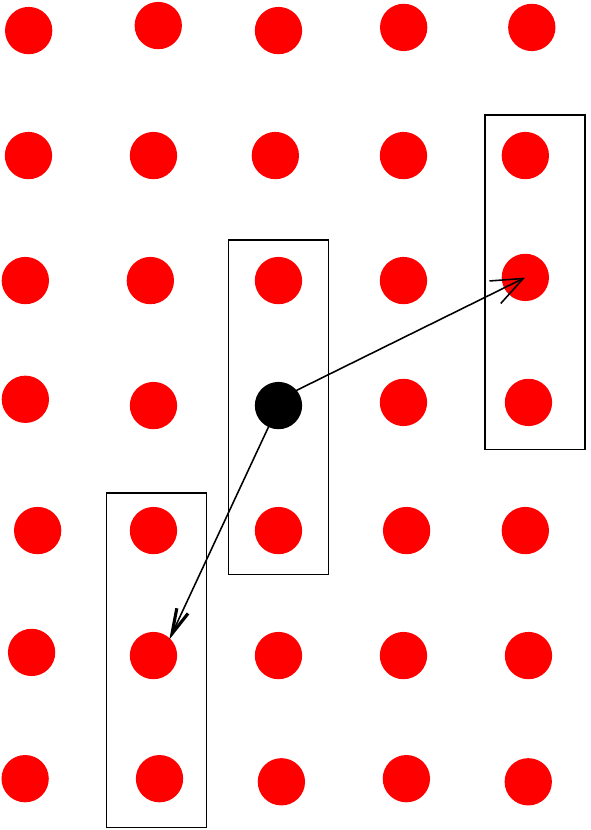}
\caption{Local configurations of 3-clusters in the NE-SW model.}
\label{fig:3clusters.stable}
\end{center}
\end{figure}

Next, we prove that the origin has a non-zero probability of participating 
in an infinite cluster for the $k=6$ NE-SW model. We divide 
the lattice into clusters of three sites, as shown in Fig.~\ref{fig:3clusters}. 
Certain clusters are defined as adjacent, and connected 
by directed lines in Fig.~\ref{fig:3clusters}. Sites not 
grouped in 3-clusters play no role in the proof, and can 
be ignored. Now, suppose that $p>(p_c^{DP})^{1/3}$,
where $p_c^{DP}$ is the critical probability for two-dimensional 
directed percolation. Then each 3-cluster is occupied with 
probability $p^3>p^{DP}_c$. The directed lines in Fig. 6 
are isomorphic to two-dimensional percolation. Thus, the 
probability of the origin participating in an infinite chain 
of 3-clusters to the northeast is nonzero, provided the 
3-cluster at the origin is occupied. Similarly, the probability 
of the origin participating in an infinite chain of 3-clusters 
to the southwest is also nonzero. Looking only at the infinite
chain of 3-clusters, it is straightforward to check that 
every site in it
is stable under the culling rules. 
To see this, observe that
each 3-cluster in the infinite path has only four possible
configurations of adjacent 3-clusters, shown in Fig.~\ref{fig:3clusters.stable}.
For all four configurations, all sites in the central 3-cluster 
are stable under the $k=6$, NE-SW stability condition.
Therefore, $p_c<(p_c^{DP})^{1/3}<1$. This bound holds for 
$k\leq 6$ for the force-balance model, since the 
force-balance constraints are a subset of the constraints 
in the NE-SW model. 

A similar proof holds for the 16 nearest-neighbor force-balance 
model. For this model, one can construct 2-clusters to obtain 
a proof that for $k\leq 4$
we have $p_c\leq (p_c^{DP})^{1/2}$. 

We note that a simple lower bound for $p_c$ for both models is the 
$p_c$ of the corresponding uncorrelated percolation models (i.e.
the $p_c$ of the model without culling).

Regarding the three-dimensional model, as was shown in 
Ref.~\cite{JammingPerc.JSP}, for three-dimensional versions 
of the jamming percolation models, percolating structures 
along one direction can cross without having sites in common. 
This prevents an obvious mapping to directed percolation. 
We have therefore not obtained a rigorous bound for $p_c$ in the 
three-dimensional model.


\subsection{Nature of the transition?}

While it can be shown that $p_c<1$ for the two-dimensional
models, there is no proof as to whether the transition is continuous or discontinuous. For the two-dimensional jamming percolation
models, there exists a rigorous argument for a discontinuous transition assuming a conjecture regarding directed percolation~\cite{Knight,Spiral.longer}. The argument 
relies on having two independent directed percolation
processes arising from the disjoint pairs of sets, each 
containing two sites. The jamming percolation transition can then be shown to occur 
at the directed percolation transition point. In the force-balance 
models considered here, the sets are composed of more than 
two sites, and are not disjoint. The critical occupation
probability thus cannot be easily identified. This means
that it is not obvious how to construct 
a rigorous argument along the lines 
of the jamming percolation models. 

However, we first argue that the alteration of some properties of the jamming percolation models to make them look more like force-balance percolation should not change the nature of the jamming percolation transition. For example, the 
 property of having two sites per set 
in the jamming percolation models can be extended to having more sites per set by investigating other
lattice models with more nearest neighbors, 
but belonging to the same universality class as  
directed percolation. Surely, there exist other directed percolation processes with more than two neighbors per site.

\begin{figure}[htb]
\begin{center}
\includegraphics[width=0.5\columnwidth]{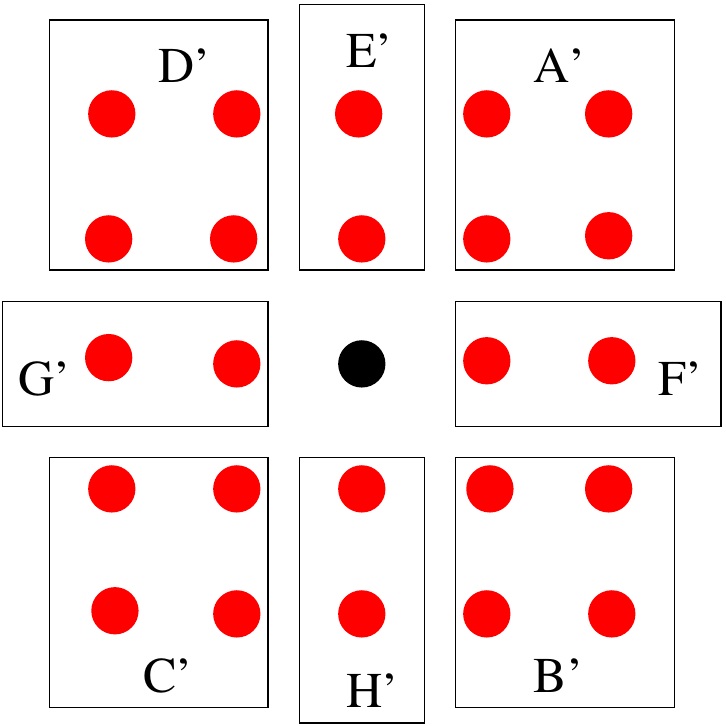}
\caption{The force-balance rules can be implemented with (A' and C') or (D' and B') or (E' and B' and C') or (F' and C' and D') or (G' and A' and B') or (H' and A' and D') or 
(A' and G' and H') or (B' and E' and G') or (C' and E' and F') or (D' and
F' and H') or (E' and F' and G' and H'). }
\label{fig:disjoint}
\end{center}
\end{figure}

On the other hand, the force-balance rules can be modified to be more like 
the jamming percolation models, where the ``and'' between
overlapping sets is equivalent to ``or'' between various
pairs or triplets of smaller disjoint sets. See
Fig.~\ref{fig:disjoint}. While the occupation of pairs of
disjoint sets are the same as in the jamming percolation case (excluding the $k$-core constraint where more than one site in the set must be occupied for the pairs), the occupation of triplets are more stringent. It remains to be seen whether or not a rigorous argument can be constructed for force-balance percolation. 

Given these arguments, the transition for the force balance percolation models may in fact behave similarly to the jamming percolation models and so we expect the transition in the force-balance case to be discontinuous as well. We will numerically test this hypothesis as well as others in the upcoming section. 
 
\section{Numerical results}
\label{sec:numerical}

 
\subsection{24 NN model}

\subsubsection{Culling dynamics}
\label{subsec:mean.cull.time}

We begin our numerical simulations of the 24 
NN (nearest neighbor)
force-balance model 
by looking at the dynamics of the culling process. 
The lattice is size $L$ by $L$.
We surround it with
a boundary of fully occupied sites, two layers thick, which
are never culled. We denote this as wired boundary
conditions. Then, at each timestep, we simultaneously cull
every unstable site. We repeat this until all remaining
sites are stable and record the culling time. 

\begin{figure}[htb]
\begin{center}
\includegraphics[width=0.9\columnwidth]{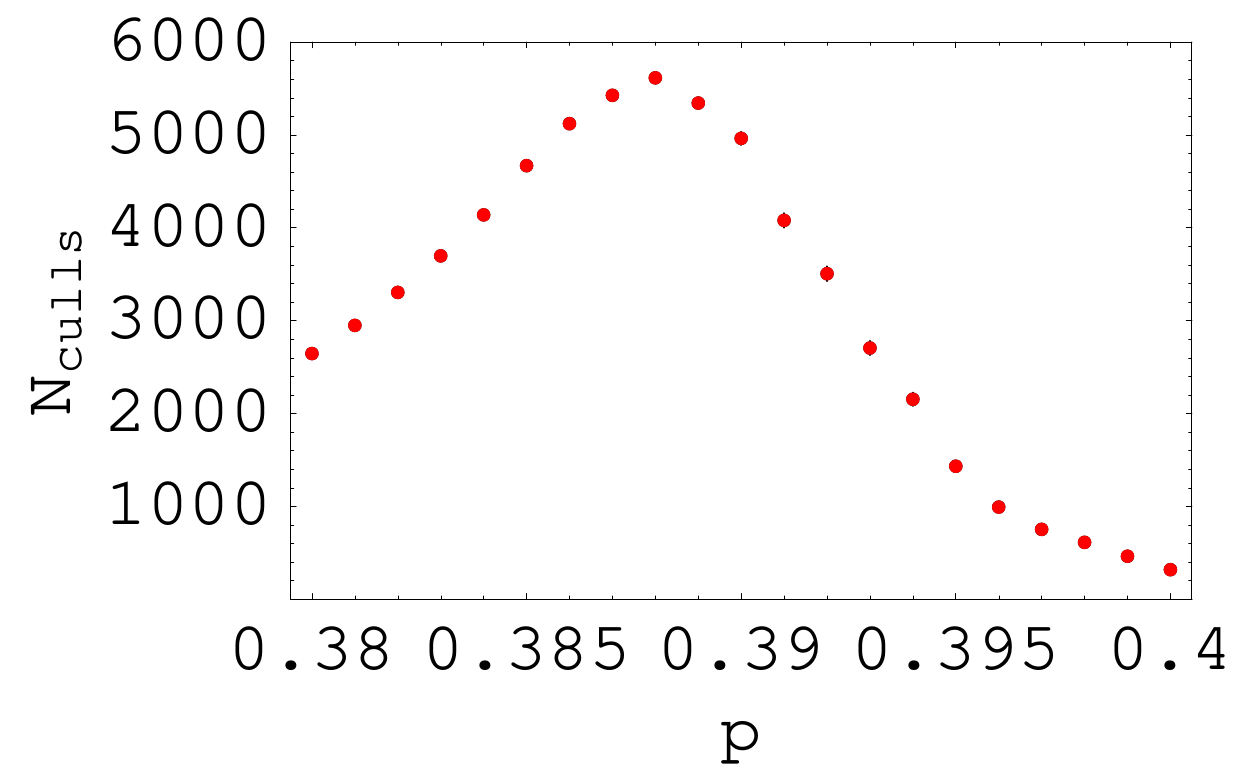}
\caption{Mean culling time as a function of initial
occupation density for the 24 NN model, $L=2896$.}
\label{fig:Z24.wired.Culling.Times.L2896}
\end{center}
\end{figure}

\begin{figure}[htb]
\begin{center}
\includegraphics[width=0.9\columnwidth]{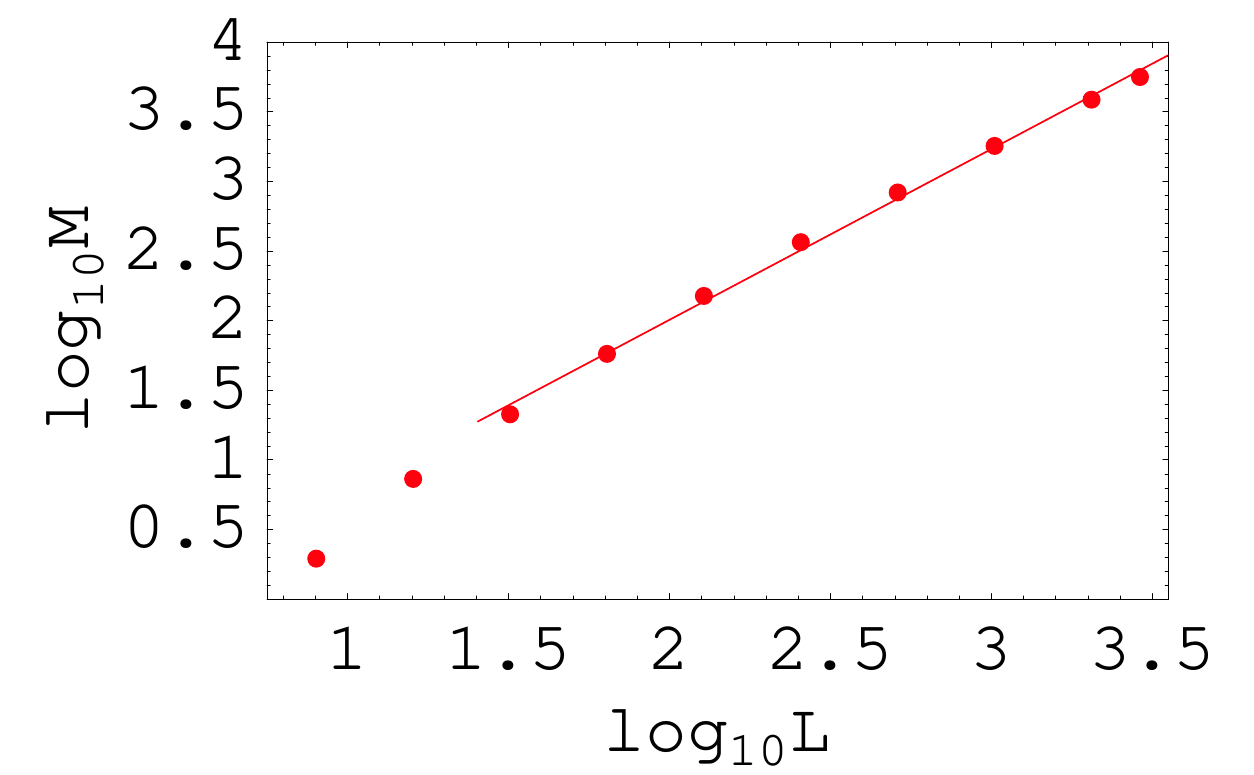}
\caption{Peak of the mean culling time as a function of $L$ for
the $24$ NN model. The best fit has a slope of
$1.226\pm0.027$. The same data is also plotted for the spiral model.}
\label{fig:Z24.wired.Culling.Time}
\end{center}
\end{figure}

For any given system size, the mean culling time as a function 
of the initial occupation density peaks for some density---a 
sample curve for $L=2896$ is shown in 
Figure~\ref{fig:Z24.wired.Culling.Times.L2896}.
The density at which the curve peaks provides one possible 
definition of the critical density. The size of the 
peak is denoted by $M$, the maximum number of average culls. $M$ grows with 
$L$. We plot $M$ vs. $L$ on a log-log scale in 
Figure~\ref{fig:Z24.wired.Culling.Time}. 
The results are well fit by a power law, although there is
visible overall curvature. A best fit for $L\geq 32$ gives $M\propto 
L^{\alpha}$, with $\alpha=1.226\pm 0.027$, which is suggestively 
close to $5/4$. 

We note that $5/4$ is the dynamical exponent for avalanches 
in the sandpile model~\cite{ASM.5over4.numerics}. Priezzhev 
has given an analytical argument that the dynamics exponent 
for the sandpile model should be exactly
$5/4$~\cite{ASM.5over4.Priezzhev}.
More generally, Pietronero, {\it et al.}~\cite{ASM.5over4.numerics} have given a renormalization
group argument that the avalanche exponent should be the 
same for a wide class of ``sandpile-like'' 
models~\cite{ASM.5over4.RG}.
They obtain an approximate exponent of $1.253\approx 5/4$.
``Sandpile-like'' models are those in which energy is dissipated, 
so that instabilities propagate from site to site.
A similar process may be responsible for the exponent of 
$5/4$ seen in our models here: culling at a site may be 
analogous to toppling at a critical point of a sandpile-like 
model, in the sense that the culling 
of one site triggers adjacent cullings that can propagate 
long distances.

\subsubsection{Force-balance avalanches}
Farrow, {\it et al.} ~\cite{Duxbury} studied the dynamics of culling 
in $k$-core percolation on various lattices by looking at avalanches. More 
precisely, the culling process is iterated until a stable $k$-core 
configuration is obtained. Then, a random site in the stable $k$-core 
is removed. It triggers the removal of other sites until the $k$-core stabilizes once more. The number of sites removed during this removal is the culling avalanche size. The process is repeated until the lattice is completely empty. 
In cases where $p_c=1$ and the stability rules allow 
no finite clusters, a power-law distribution of culling 
avalanche sizes was found. For these cases, 
the system goes from being unoccupied for $p_c<1$, to being 
fully occupied at $p_c=1$. Since all sites must be eventually be removed below $p_c=1$ and no finite clusters are allowed, the culling avalanches should become spatially long-ranged near $p_c=1$. This result is to be contrasted 
with $k$-core cases with finite $k$-core clusters. There, 
no power law distribution of culling avalanche sizes was 
found~\cite{Duxbury}. 

In the force-balance model, there 
are no finite force-balance clusters, so the culling avalanches
should be spatially long-ranged near the transition. Whether or not there should be a broad distribution of sizes is not clear a priori. Given the sandpile-like behavior detected in the mean culling time required to obtain a force-balance cluster, one may expect to find a broad distribution as is found in sandpile models.  However, if 
the force-balance percolation transition is
discontinuous one would expect a well-defined avalanche size
for those systems whose redundant sites have already been
removed. 

\begin{figure}[htb]
\begin{center}
\includegraphics[width=0.9\columnwidth]{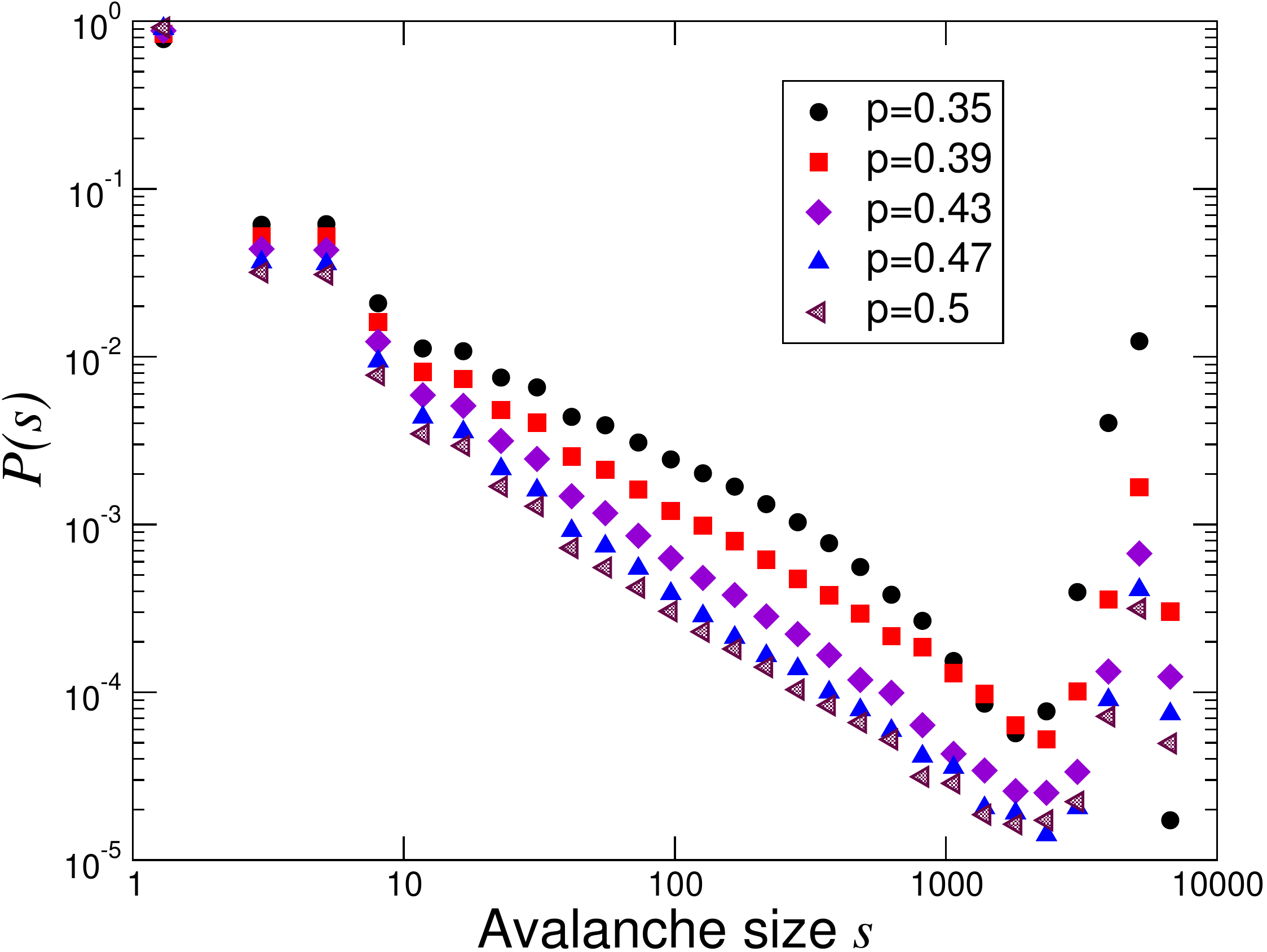}
\caption{Log-log plot of $P(s)$ for $L=128$ using periodic boundary conditions.}
\label{fig:avalanche1}
\end{center}
\end{figure}

Numerically, we calculate the probability of having a culling avalanche size $s$, $P(s)$ in the presence of periodic boundary conditions. See Fig.~\ref{fig:avalanche1}. The probability is broad near and above the transition for intermediate avalanche sizes. On a log-log plot, the slope of $P(s)$ for the broadly distributed intermediate-sized avalanches depends somewhat on $p$ and on $L$. More careful study is needed to determine whether or not these avalanche sizes are consistent with the measured $-1.253$ for sandpile models~\cite{ASM.5over4.RG}. It is of note that for $p=0.45$ and $L=512$, for example, the slope is approximately $-1.3$. 

\begin{figure}[htb]
\begin{center}
\includegraphics[width=0.9\columnwidth]{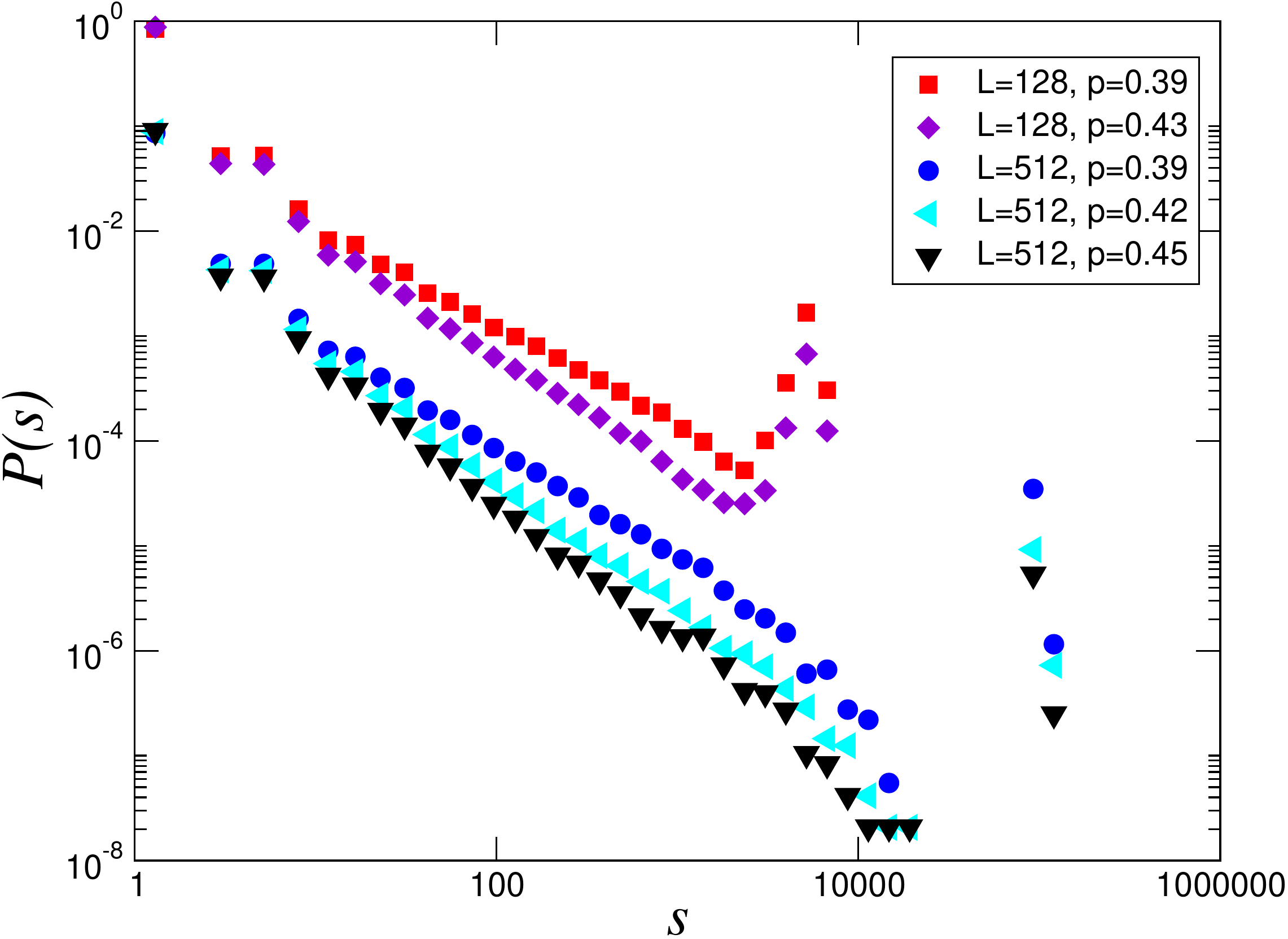}
\caption{Log-log plot of $P(s)$ for $L=128$ and $L=512$ for comparison. The $L=512$ curves have been shifted downward by a factor of 0.1.}
\label{fig:avalanche2}
\end{center}
\end{figure}

As opposed to quantitative analysis of the intermediate-sized avalanches, we are more interested in first determining the qualitative nature of the curves. As opposed to a purely broad distribution, there is a prominent peak at the tail of the distribution. These well-defined avalanche sizes correspond the last occuring avalanches when the lattice ultimately empties out. These correspond to the marginal infinite cluster. The peak persists as $p$ is decreased towards the $p_c$ indicated from the position of the maximum in the mean culling times, suggesting a 
discontinuous transition for the onset of the infinite cluster. This behaviour is retained in the larger systems with the peak becoming more separated from the broad part of the distribution. See Fig.
~\ref{fig:avalanche2}. 

\subsubsection{Spanning cluster}

\begin{figure}[htb]
\begin{center}
\includegraphics[width=0.9\columnwidth]{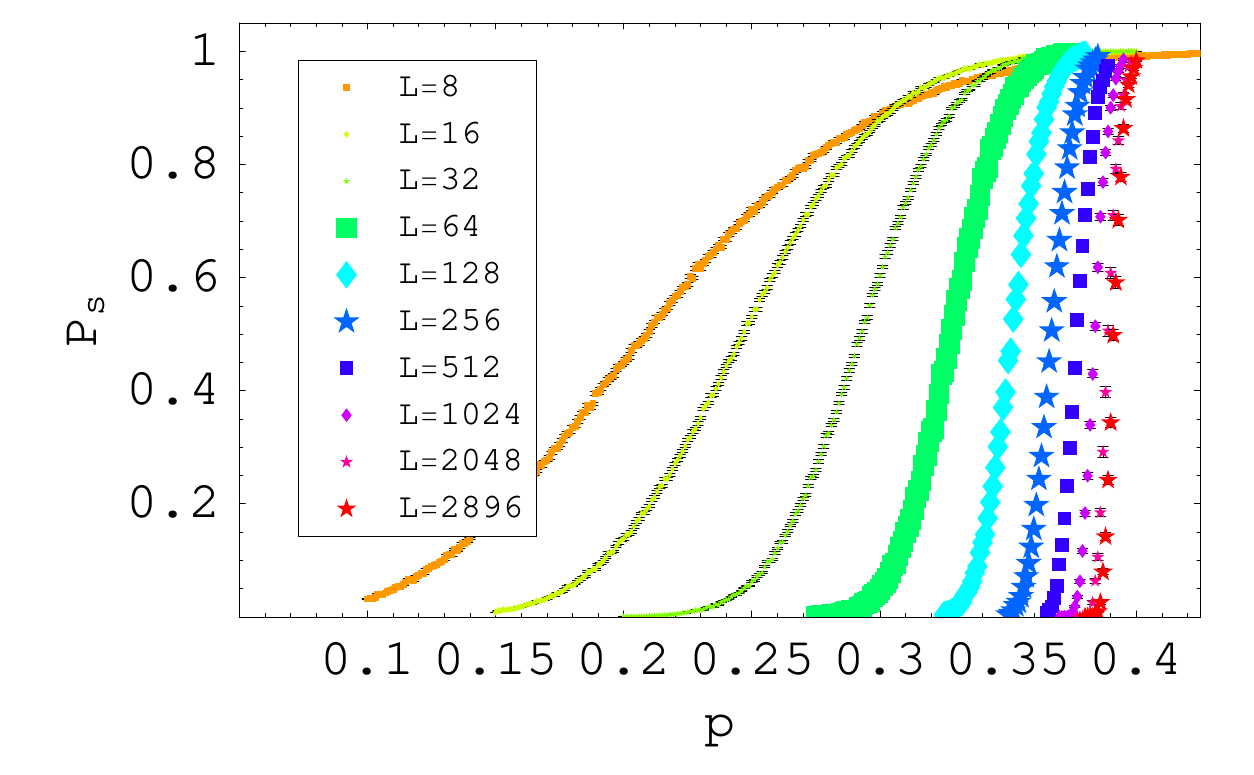}
\caption{The probability of spanning for wired boundary
conditions in the $24$ NN model.}
\label{fig:Z24.wired.Pspan}
\end{center}
\end{figure}

We next look at $P_s$, the probability of spanning.
We define this, for wired boundary conditions, as the
probability that the largest cluster connects either the top 
and bottom sides, or the right and left sides; for this
definition, sites are defined as connected if one is within 
the 24 site neighborhood of the other
(see Fig.~\ref{fig:z24.definition}). 
Fig.~\ref{fig:Z24.wired.Pspan}
depicts $P_s$ for the 24 NN model with wired boundary conditions. 

\begin{figure}[htb]
\begin{center}
\includegraphics[width=0.9\columnwidth]{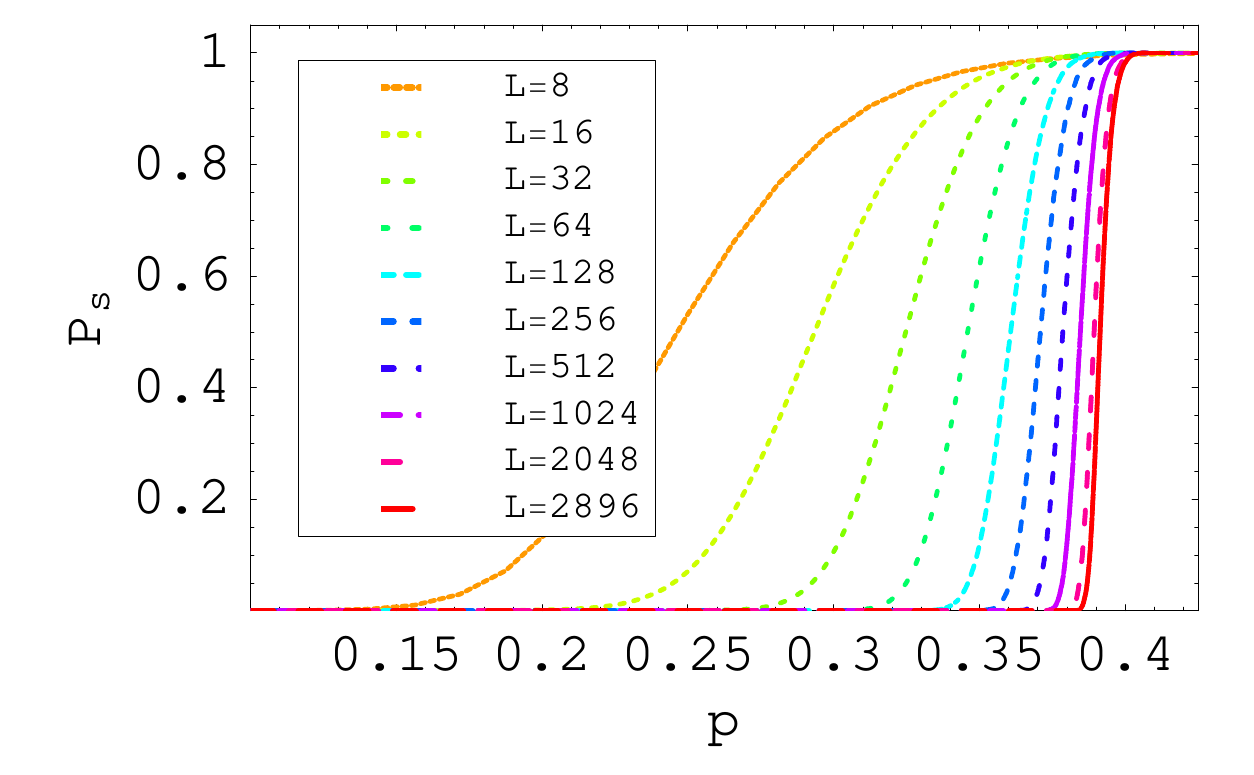}
\caption{The probability of spanning for periodic boundary
conditions in the $24$ NN model.
Error bars are too small to be seen.}
\label{fig:Z24.periodic.Pspan}
\end{center}
\end{figure}

Fig.~\ref{fig:Z24.periodic.Pspan} shows the same curves, 
but for periodic boundary conditions. For periodic boundary 
conditions, it is impossible to have any occupied sites 
without having a spanning cluster, so to check if we have 
a spanning cluster, we just need to check that at least 
one site is occupied.

For periodic boundary conditions, the $P_s$ curves can be 
generated quickly by a simple modification of the
Newman-Ziff algorithm~\cite{Newman.Ziff}. 
For uncorrelated percolation, the Newman-Ziff algorithm passes 
once through every possible density, in increasing order,
by adding occupied sites in random order, and updating the 
connectivity with the Hoshen-Kopelman algorithm~\cite{Hoshen.Kopelmann}.
This allows for efficient generation of $P_s$ for every 
occupation density. This method is not available to us for 
the force-balance model with wired boundary conditions, 
because there sites are added by increasing the density, 
but also removed by culling. This prevents a single pass 
through all densities from being done in time
$\mathcal{O}(L^2\ln L)$. 
(The Hoshen-Kopelmann algorithm does not allow cluster identifications 
to be rapidly updated after removal of sites.)
However, with periodic boundary conditions in a force-balance 
model, we do not need to run the Hoshen-Kopelman algorithm. 
We can instead start from a fully occupied lattice, and 
remove sites until the culling condition causes an empty 
lattice. This allows the calculation of $P_s$ at every density 
in time $\mathcal{O}(L^2)$.

\begin{figure}[htb]
\begin{center}
\includegraphics[width=0.9\columnwidth]{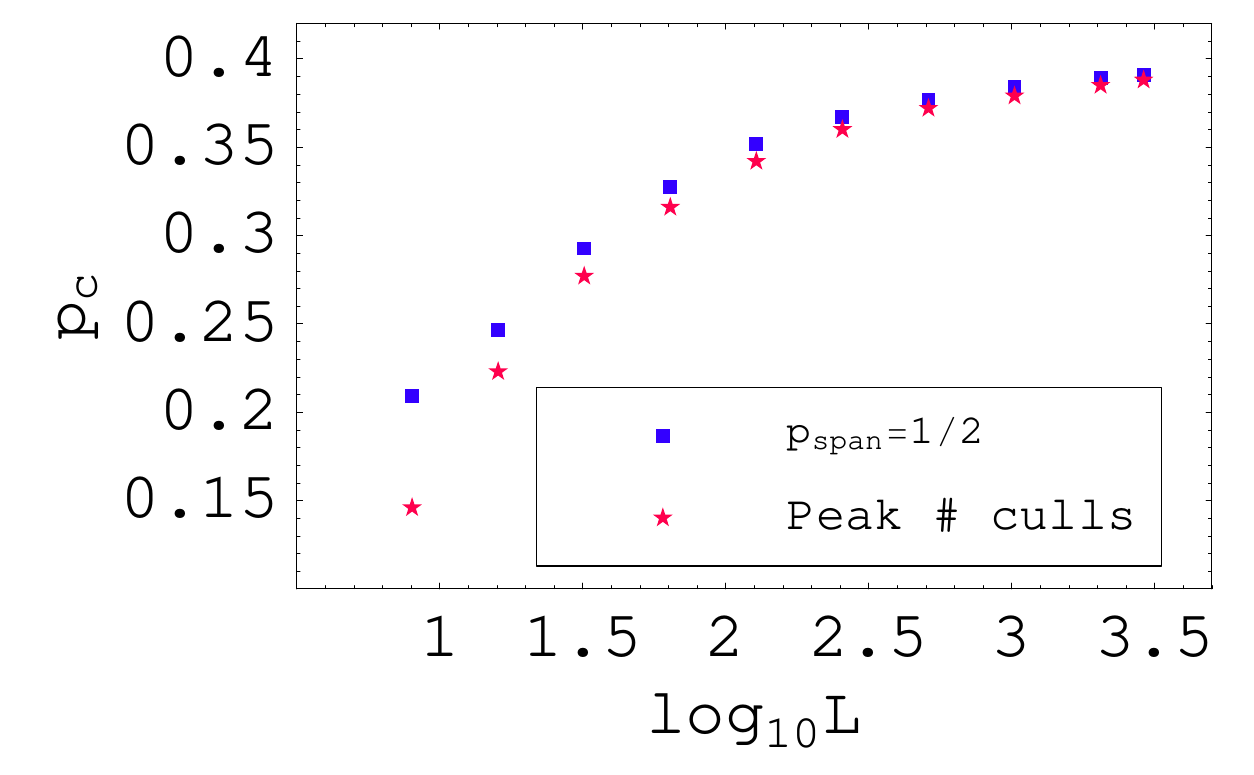}
\caption{$p_c(L)$ for the $24$ NN model, using two
definitions of the critical point: (1) the density at which
$P_s=1/2$, and (2) the density at which the mean
culling time peaks.}
\label{fig:Z24.wired.PCrit.TwoMethods}
\end{center}
\end{figure}

From Figs.~\ref{fig:Z24.wired.Pspan} and~\ref{fig:Z24.periodic.Pspan}, 
we can define the critical probability for a specified system 
size, $p_c(L)$, as the initial occupation density that gives 
$P_s=1/2$. This definition differs from that based on the 
peak of the mean culling time, 
given in subsection~\ref{subsec:mean.cull.time}. However, 
as shown in figure~\ref{fig:Z24.wired.PCrit.TwoMethods}, 
the two definitions have the same qualitative dependence 
on $L$, and approach each other for large system sizes.

Given $p_c(L)$ we can extract the thermodynamic critical 
point, $p_c(L=\infty)$, if given the functional 
dependence of $p_c(L)$ on $L$. However, it is unclear what 
functional dependence to fit to.
It was assumed in Ref.~\cite{SLC} that 
$|p_c(L=\infty)-p_c(L)|\propto L^{-1/\nu}$, where $\nu$ is an exponent characterizing the
divergence of the length scale.
With this functional 
form, $\nu$ can be detected by varying
$p_c(\infty)$ until a log-log plot gives a straight line.
However, such a procedure is not a check of the functional 
form, since we are allowed to choose $p_c(\infty)$.

\begin{figure}[htb]
\begin{center}
\includegraphics[width=0.9\columnwidth]{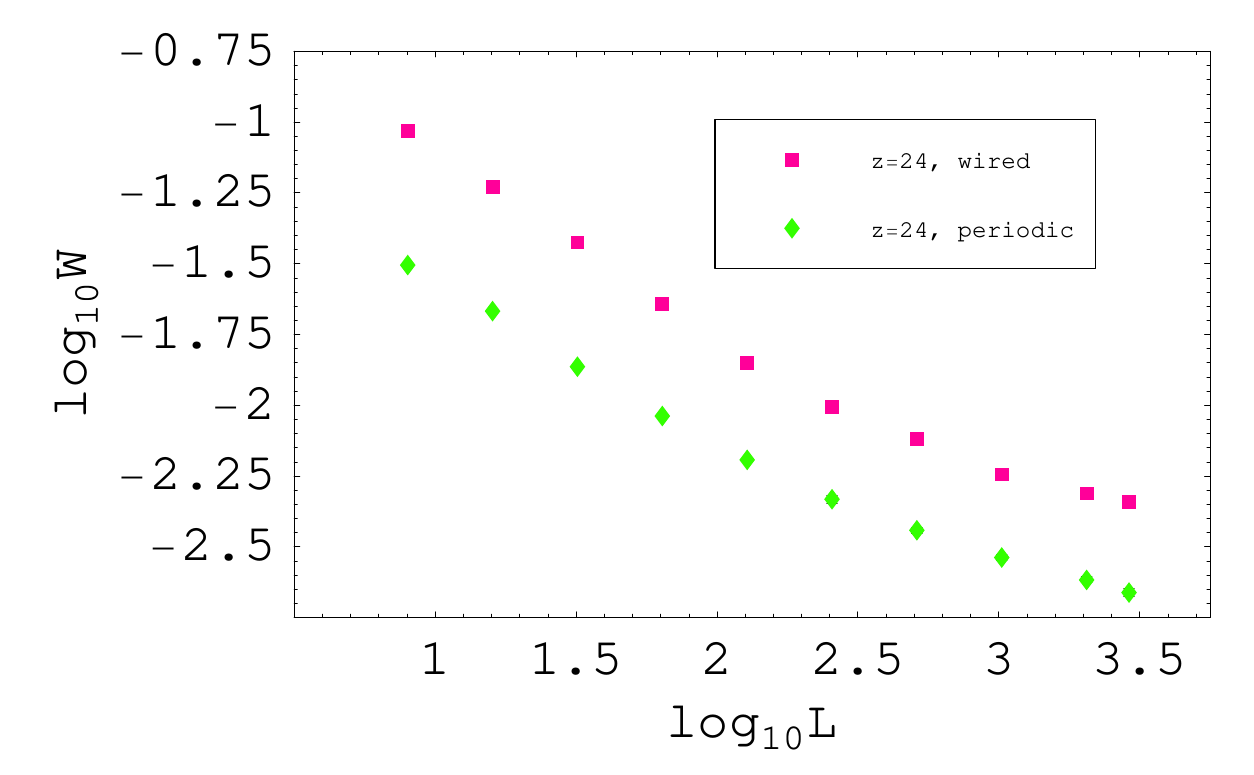} 
\caption{The widths for the 24NN model with both wired and
periodic boundary condition, on a log-log plot. Both boundary conditions 
show clear deviations from power
laws at large system sizes.}
\label{fig:Z24.Widths.Plot1}
\end{center}
\end{figure}

A check of the functional form can be done by looking at 
the width $W$ of the transition, which we define to be the 
difference between the occupation probabilities that yield 
$P_s=1/4$ and $P_s=3/4$. The width as 
a function of $L$ is shown in figure~\ref{fig:Z24.Widths.Plot1} for both wired and 
periodic boundary conditions. 
 The same renormalization group formalism that implies
$\mid p_c(L=\infty)-p_c(L)\mid \propto L^{-1/\nu}$
also implies $W\propto L^{-1/\nu}$. However,
Fig.~\ref{fig:Z24.Widths.Plot1} shows clear deviations from
this power law form. 
This trend was starting to emerge in Ref.~\cite{SLC}, though it was masked by the large error bars for the larger systems. In other words, the fit to the above assumption in Ref.~\cite{SLC} 
was premature. 

\begin{figure}[htb]
\begin{center}
\includegraphics[width=0.9\columnwidth]{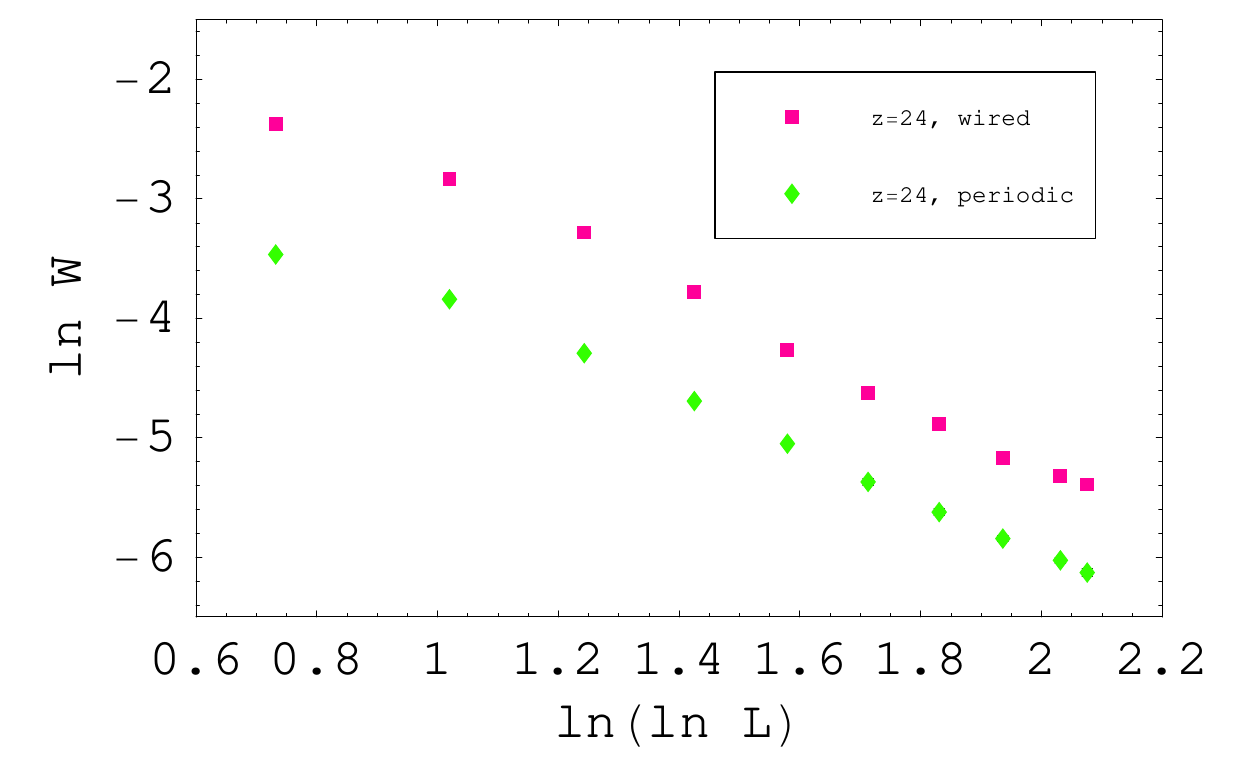}
\caption{The widths for the 24NN model with both wired and
periodic boundary condition, using a fitting form motivated
by the TBF results.}
\label{fig:Z24.Widths.Plot2}
\end{center}
\end{figure}

Based on our above heuristic arguments for a connection between force-balance percolation and jamming percolation, we instead fit 
to the form for a diverging crossover length scale 
found by
Toninelli, {\it et al.}~\cite{TBF.response,Spiral.longer} (TBF) for the spiral model. Approaching the transition from below, TBF 
identified a crossover length $\Sigma$, such that systems 
of size smaller than $\Sigma$ are likely 
to have a spanning cluster, while those of size greater 
than $\Sigma$ are exponentially unlikely to have a spanning 
cluster. TBF proved that $\Sigma$ diverges at the transition 
faster than any power law. In Ref.~\cite{Knight}, TBF argued that the upper and lower bounds scaled similarly, finding
$\Sigma\sim \exp(-C(p-p_c)^{-\mu})$, where $\mu=(1-\frac{1}{z})\nu_{||}\cong 0.64$. 
(Both $\frac{1}{z}\cong 0.63$ and $\nu_{||}\cong 1.73$ are
from directed percolation~\cite{DP}.) Based on the TBF formula,  
we propose that $W\propto (\ln(L))^{-1/\mu}$. 
The widths are replotted with the appropriate scales for
this fit in Figure~\ref{fig:Z24.Widths.Plot2},
and while the curves still have deviations from perfect
straight line behavior, the agreement is significantly better
than for $W\propto L^{-1/\nu}$. From the width data, for wired boundary conditions, we extract $\mu=0.39\pm0.01$; for periodic boundary conditions, $\mu=0.45\pm0.02$. The stated error bar is purely statistical and does not take into account systematic effects that occur given the small range of system sizes, so one cannot necessarily rule out a link with overlapping directed percolation processes. We also note that in Ref.~\cite{Knight}, TBF stated that their numerical data was consistent with $\mu\cong0.64$ given the large systematic error bars.

\begin{figure}[htb]
\begin{center}
\includegraphics[width=0.9\columnwidth]{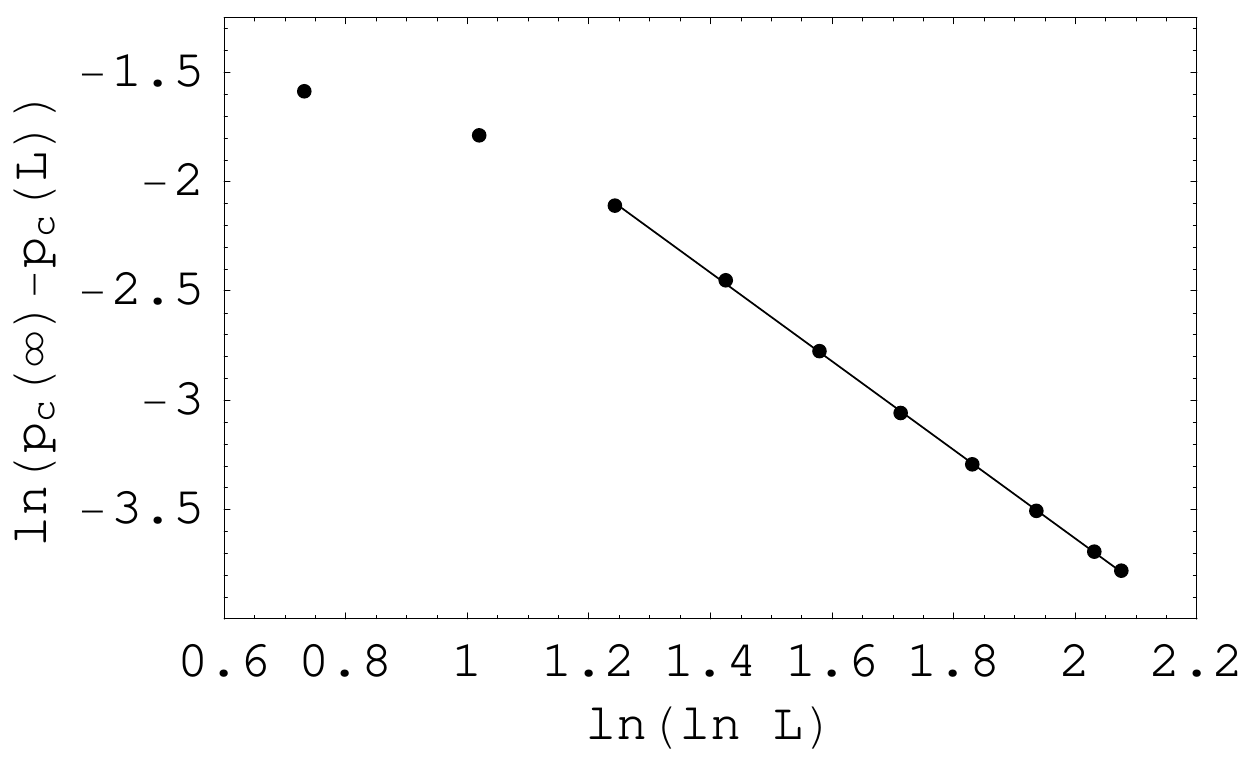}
\caption{Fitting of $p_c(L)$ as a function of $L$ using the
TBF fitting form, for the $24$ NN force-balance model with wired boundary conditions. 
We obtain the best fit with
$p_c(\infty)=0.414\pm0.008$.}
\label{fig:Z24.wired.PCrit.TBFfit}
\end{center}
\end{figure}

Now, if there is only 
one diverging lengthscale, then it can be argued that $|p_c(L)-p_c(\infty)|$ should be proportional to $W$. So we choose the TBF fitting form 
and choose $p_c(\infty)$ to minimize $\chi^2$ for 
the best linear fit of $\ln (p_c(\infty)-p_c(L))$ vs. 
$\ln (\ln L)$, for $L\geq 32$.
The optimal fit, shown in
figure~\ref{fig:Z24.wired.PCrit.TBFfit},
is obtained for $p_c(\infty)=0.414\pm 0.008$ for wired boundary conditions.
(The error bar is obtained by testing which values of
$p_c(\infty)$ result in visibly noticeable curvature in the 
plot.) For periodic boundary conditions, we obtain $p_c(\infty)=0.425\pm0.005$. 
The straight line obtained in
figure~\ref{fig:Z24.wired.PCrit.TBFfit} does not make 
clear whether the TBF form is correct for our
model, given that we have the freedom to choose
$p_c(\infty)$ in making the fit. From the fit with wired boundary conditions we obtain $\mu=0.51\pm 0.09$. Again, this result is to be compared with TBF's result of $\mu=0.64$. 

From the two different results for $\mu$,  
it appears that $W$ and $|p_c(L=\infty)-p_c(L)|$ do not
scale in the same way. There are two possible, but incompatible, explanations 
for the difference in the results for $\mu$. One possibility is that
the scaling of $W$ and $|p_c(L=\infty)-p_c(L)|$ 
is in fact the same in the thermodynamic limit, but that we
obtain differing $\mu$'s for the finite system sizes studied
due to the slow approach to the thermodynamic limit. This
would be reminiscent of the extremely large finite-size
effects seen in analogous systems, such as $k$-core
percolation, the Kob-Andersen model, and glassy dynamics.
Numerically, this possibility is quite reasonable, given
that the TBF formula for the growing length scale has an
extra power law factor in front. A second possible
explanation is that $W$ and 
$|p_c(L=\infty)-p_c(L)|$ in fact scale differently. This
would be possible if there are are two diverging length
scales, rather than one; this would change the standard
picture of a renormalization group fixed point controlled by
a single parameter, so that 
$W$ and $|p_c(L=\infty)-p_c(L)|$ 
would no longer be forced to scale in the same manner. There are certainly two diverging lengthscales in mean field $k$-core percolation---one associated with those in the infinite cluster and one associated with sites that get removed in response to one random site being removed.

\subsubsection{Order parameter}

\begin{figure}[htb]
\begin{center}
\includegraphics[width=0.9\columnwidth]{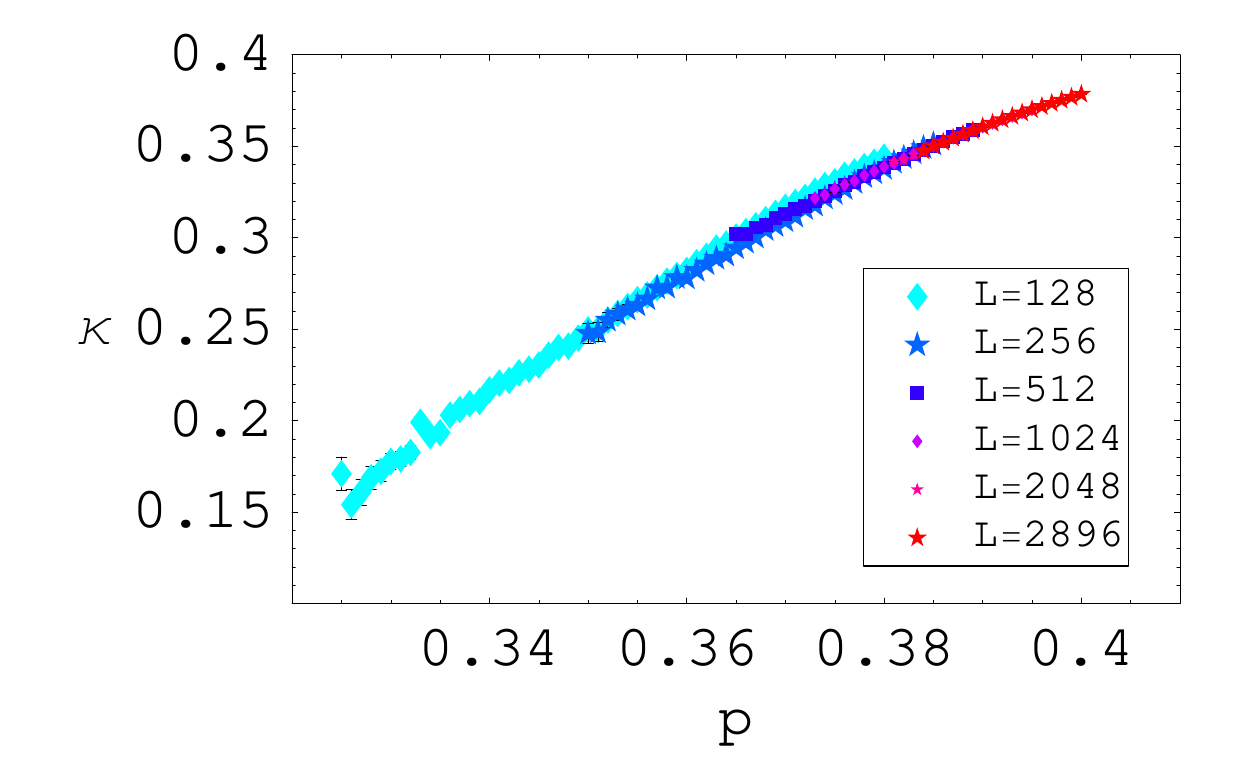}
\caption{Order parameter for the $24$ NN model using wired boundary conditions.}
\label{fig:Z24.wired.OrderParameter}
\end{center}
\end{figure}

We also investigate the order parameter, $\kappa$,
the fraction of sites in the infinite force-balance cluster.
This is obtained for a given system size and initial occupation 
density by generating configurations, keeping only those 
configurations in which the largest cluster is spanning, 
and then finding the average fraction of sites in the largest 
cluster for those configurations. Plots are shown for $L\geq 
128$ in figure~\ref{fig:Z24.wired.OrderParameter}.
The curves lie on top of each other for large $L$, differing 
only in the minimum value of $p$ needed 
for the numerical simulations to have a reasonable 
chance of obtaining occupied sites after culling.

The minimum value of $\kappa$ is increasing, rather than
decreasing, with system size. We therefore assume that 
there is a jump in the order parameter at the transition,
denoted by $\kappa_c$. In other words, in the thermodynamic
limit, as soon as there is a spanning cluster, it occupies a
finite fraction of the system.

From Fig.~\ref{fig:Z24.wired.OrderParameter},
and the already-obtained result of $p_c=0.414\pm 0.008$,
we estimate the thermodynamic $\kappa_c$ 
to be $\kappa_c =0.399\pm 0.011$. 
Furthermore, for each individual curve 
it appears that just above the transition, $\kappa$ increases 
linearly with $p$, suggesting that the order parameter exponent
is $\beta=1$. This result is to contrasted with the mean field 
$k$-core percolation results and the repulsive soft sphere 
simulations, where $\beta=1/2$.

\subsubsection{Correlation function}

We next look at correlation functions for this model, 
now using only periodic boundary conditions. For these 
simulations, for a given system of size $L$ and density 
$p$, we generate
states with exactly $pL^2$ occupied sites (rather than occupying 
sites independently). We only keep configurations that 
were nonempty after culling, and in each of these, keep 
only the largest cluster (this cluster was automatically 
spanning, since we use periodic boundary conditions).
We then calculate the correlation function $C(i)$---the 
probability that sites a distance $i$ apart are both occupied, 
minus an asymptotic constant, to be discussed shortly. 
Moreover, we calculate correlations at all angles (i.e. not merely
horizontal and vertical correlations).

With uncorrelated percolation, it is easy to quickly produce precise 
plots of the correlation function over several
decades of distance, and confirm that the correlation function 
has a power law form at the critical point. However, correlations 
in the force-balance model have significantly greater sources 
of uncertainty.

 When we generate samples
for uncorrelated percolation, each site occupation is independent, 
so the correlation function calculated for $C(i)$ is largely 
(although not entirely) independent of that for $C(j)$ 
with $j\neq i$. So for uncorrelated percolation, when we look 
at different pairs of points in a single sample, 
we get somewhat independent estimates of the correlation 
function for a range of distances. That is, we quickly generate 
lots of independent data for the curve $C(i)$. However, 
for correlated percolation models where there are no finite 
clusters, the results for $C(i)$ and $C(j)$, for $j\neq 
i$, are are highly correlated for a given sample, even for 
widely separated $i$ and $j$. 

For a given amount of computer running time, the error bars
in the correlation function are thus substantially 
larger than for uncorrelated
percolation, and because they are correlated, it is
difficult to reliably extract the asymptotic value of
$C(i)$. This is important because to get $C(i)$, we need to
subtract off the probability that two 
distant sites ($i\to\infty$) are occupied in the {\it infinite system
limit} ($L\to\infty$) after culling. For our models, we do not know the infinite
system occupation probability after culling. 
(For uncorrelated percolation, it is
trivially identical to
the initial occupation probability,
as there is no culling). The most straightforward
solution is to subtract off the asymptotic value of $C(i)$,
since we know that the connected 
correlation function should
approach $0$ as $i\to\infty$.
However, because the error bars at different $i$ are
correlated, there are also large error bars in the
asymptotic value of $C(i)$. The functional form of $C(i)$ is
highly sensitive to the constant subtracted off---over a
limited distance range, a power law and an exponential can
look very similar if the asymptotic value is shifted
slightly. 
This means that that it is difficult to
reliably determine the form of the correlation function.
The correlation length,

\begin{equation}
\label{eq:corrlength.sum}
\xi^2 = \frac{\sum_i i^2 C(i)}{\sum_i C(i)}\ ,
\end{equation}

\noindent is also very sensitive to a small shift in $C(i)$.

\begin{figure}[htb]
\begin{center}
\includegraphics[width=0.9\columnwidth]{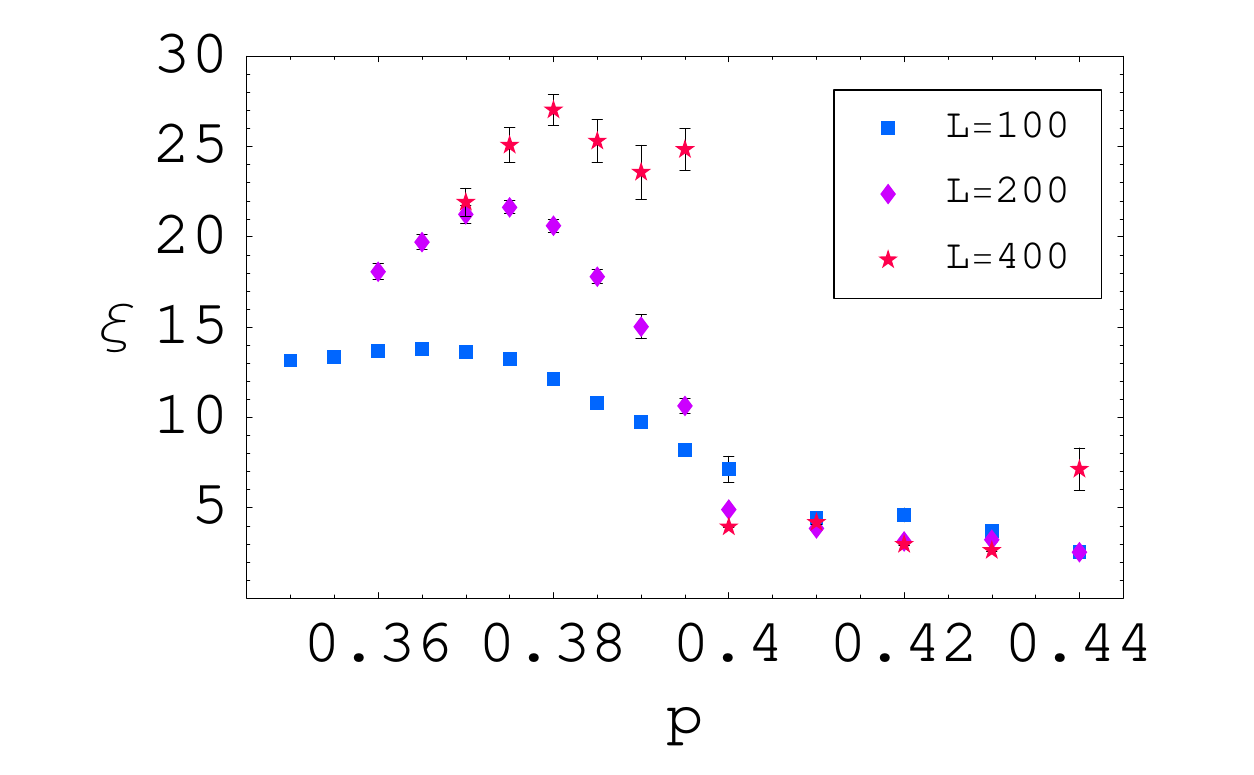}
\caption{Correlation lengths for the $24$ NN force-balance
model as a function of $p$, for $L=100$ and $L=200$.}
\label{fig:Z24.CorrelationLengths}
\end{center}
\end{figure}

The end result is that it is difficult to get accurate
results for large system sizes, and we have limited our
simulations of correlation functions to $L=100$, $L=200$ and $L=400$.
The correlation length as a function of initial occupation
density is shown for these three system sizes in 
figure~\ref{fig:Z24.CorrelationLengths}.
For densities $\rho>0.4$, the correlations are clearly
short-ranged, and are independent of the system size.
For lower densities, the correlation length grows with $L$.
The curve for $L=200$ has a clear peak at $p\approx
0.375$; the $L=400$ data appears to have a peak at a larger $L$, although the
largish error bars make its location unclear. Thus, 
the data is somewhat suggestive of a lengthscale that grows 
with system size. 

The position of the peak gives yet another plausible 
measurement for $p_c(L)$. If we compare the $p_c(L)$ defined from the probability of spanning for $L=200$, $p_c(L)=0.3671\pm0.0007$, which is lower than the estimated $p_c(L)$ from correlation length data. However, this discrepancy should vanish in the infinite system limit.  Since the position of the peak is not so clear for $L=400$ it is more difficult to discern the trend. 

\begin{figure}[htb]
\begin{center}
\includegraphics[width=0.9\columnwidth]{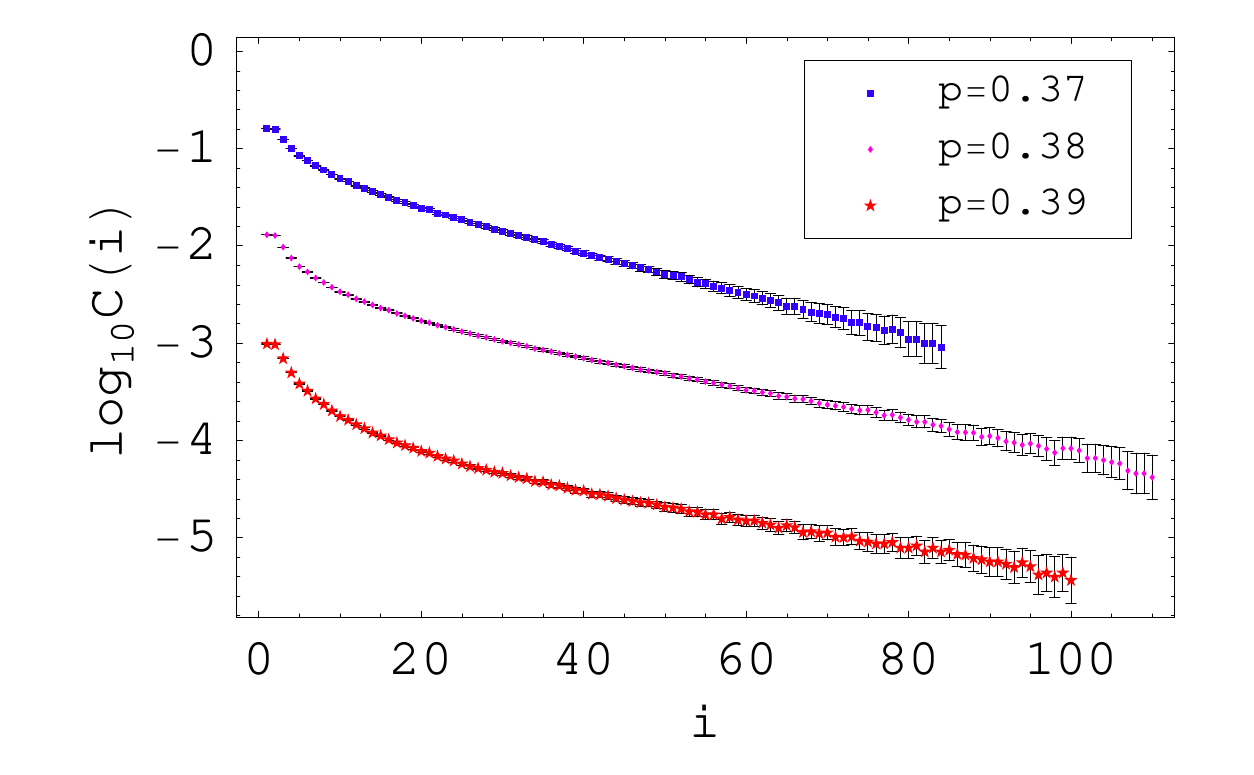}
\caption{Correlation function 
for $L=400$, for the $24$ NN force-balance model, at
$p=0.37$, $p=0.38$, and $p=0.39$.
For clarity, the $p=0.37$ correlation function has been
shifted up by 1.0, and the $p=0.39$ correlation function has
been shifted down by 1.0.}
\label{fig:Z24.CorrelationFunction.L400.multiple}.
\end{center}
\end{figure}

\begin{figure}[htb]
\begin{center}
\includegraphics[width=0.9\columnwidth]{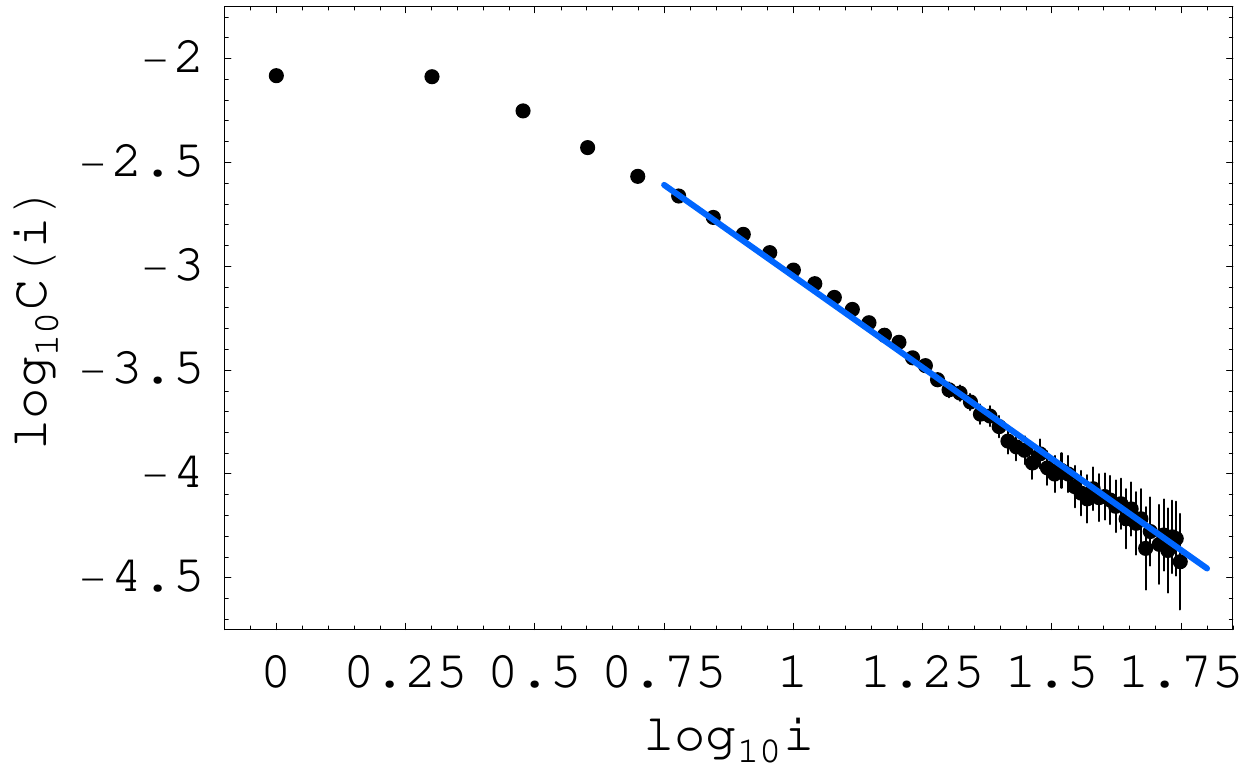}
\caption{Correlation function for $L=200$, $p=0.395$,
for the $24$ NN force-balance model. The
best fit line has a slope of $-1.76\pm0.03$.}
\label{fig:Z24.CorrelationFunction.L200.p39500}
\end{center}
\end{figure}

Because of the limited system sizes studied, it is somewhat 
difficult to determine with confidence whether the correlation 
functions have power law or exponential forms. The correlations 
are very short-ranged ($\xi < 5$) for $p\geq 0.4$, indicating 
that the correlations are probably exponential (or a very 
steep power law). At lower densities, for most system 
sizes and densities, the correlation function is exponentially 
decaying at long distances; for example, see
figure~\ref{fig:Z24.CorrelationFunction.L400.multiple}, 
showing the correlation functions for $L=400$ and $p=0.37$, 
$0.38$, and $0.39$. One exception to the generally observed 
exponentialy behavior is te $L=200$ correlation function 
for $p=0.395$, which appears somewhat power-law-like, as 
shown in figure~\ref{fig:Z24.CorrelationFunction.L200.p39500}.
The density of $p=0.395$ differs slightly from the peak 
in the plot of $\xi$ for $L=200$ in
figure~\ref{fig:Z24.CorrelationLengths}. 
For the $L=400$ system, the 
power-law-like trend of the correlation function near the 
peak density is not as prominent and the correlation functions 
appear more exponential. 

These results should be compared with those of Parisi and Rizzo 
for k-core percolation in four dimensions~\cite{Parisi}. 
While they found that $\xi$ grew with decreasing $\rho$, 
they never obtained correlation lengths greater than 10.  So they also found no power law correlations.  They argue that the 4-core transition in four dimensions is an ordinary discontinuous transition with no diverging lengthscales. We note that their systems had finite 
stable clusters and so may have different properties than 
the systems considered in this paper. (The fact that their 
systems had finite stable clusters also presumably means 
that they had smaller numerical correlations between $C(i)$ 
and $C(j)$ for $\mid i-j\mid$ large, allowing them to 
determine correlation functions for
larger systems than here.)



\subsection{16 NN Model and Spiral Model}

\begin{figure}[htb]
\begin{center}
\includegraphics[width=0.9\columnwidth]{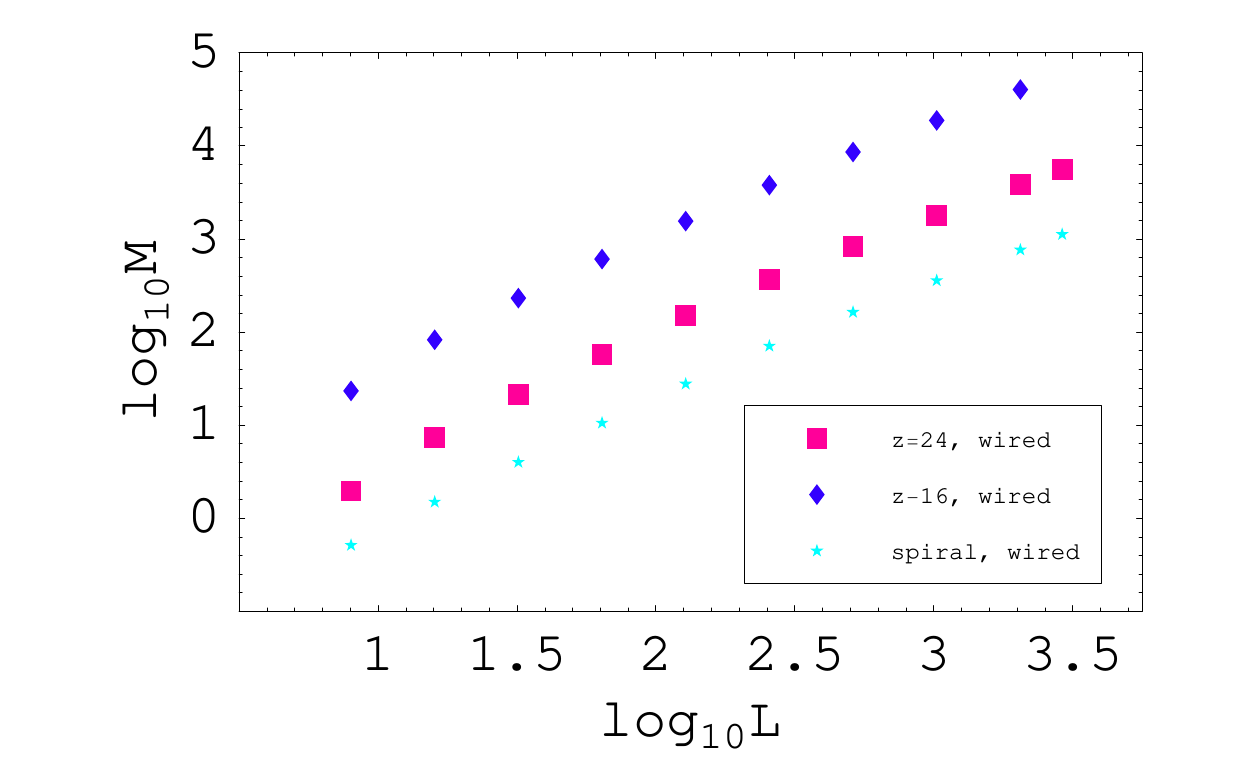}
\caption{Peak of the mean culling time as a function of $L$ for
the all three models. For the 16NN model, the best fit has a slope of
$1.240\pm0.025$. For the spiral model, the best fit has a slope of
$1.246\pm0.027$. Wired boundary conditions are used.}
\label{fig:Z16.wired.Culling.Time}
\end{center}
\end{figure}

Now we present results for the 
16 nearest neighbor force-balance model and spiral model. We begin 
with the mean culling time. 
The plot of the mean peak 
culling time as a function of $L$ shows a similar $5/4$ 
exponent as was measured for the 24NN model, $\alpha=1.240\pm0.025$. In fact, all three two-dimensional models---the 24NN model, the 16NN model and the spiral model---are in quantitative agreement with another. We find 
$\alpha=1.246\pm0.027$ for the spiral model. 
These results suggest that similar processes underlie the culling in all three 
two-dimensional models. See Fig.~\ref{fig:Z16.wired.Culling.Time}. 

\begin{figure}[htb]
\begin{center}
\includegraphics[width=0.9\columnwidth]{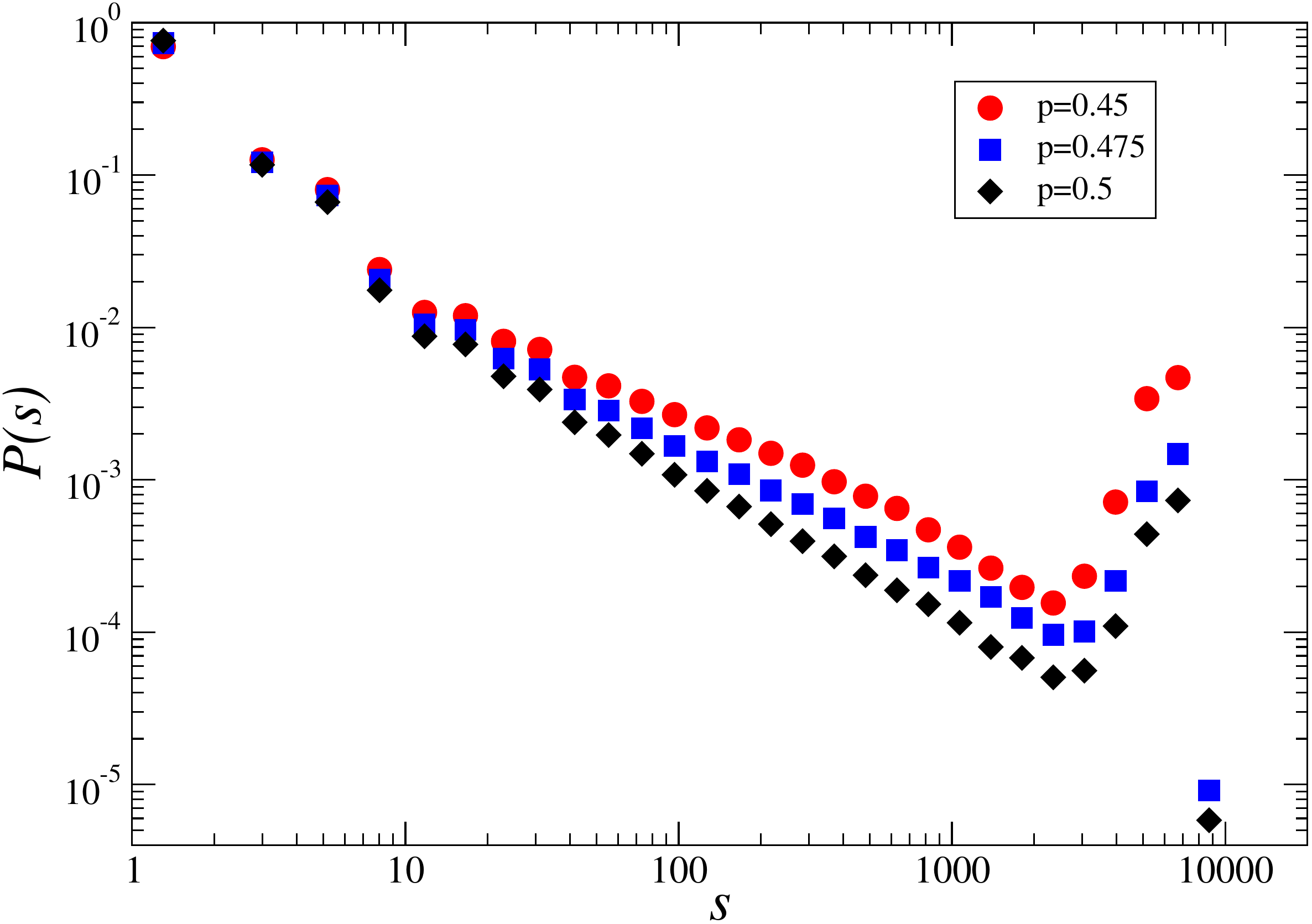}
\caption{Log-log plot of $P(s)$ for the 16 NN model with $L=128$ in the presence of periodic boundary conditions.}
\label{fig:avalanche.16NN}
\end{center}
\end{figure}

The probability of having a force-balance avalanche of size $s$, $P(s)$, for the 16 NN model shows similar trends to the 24 NN model with a broad distribution of intermediate sizes and a prominent peak for the largest sizes. See Fig.~\ref{fig:avalanche.16NN}. The spiral model exhibits the same qualitative behaviour as well.

\begin{figure}[htb]
\begin{center}
\includegraphics[width=0.9\columnwidth]{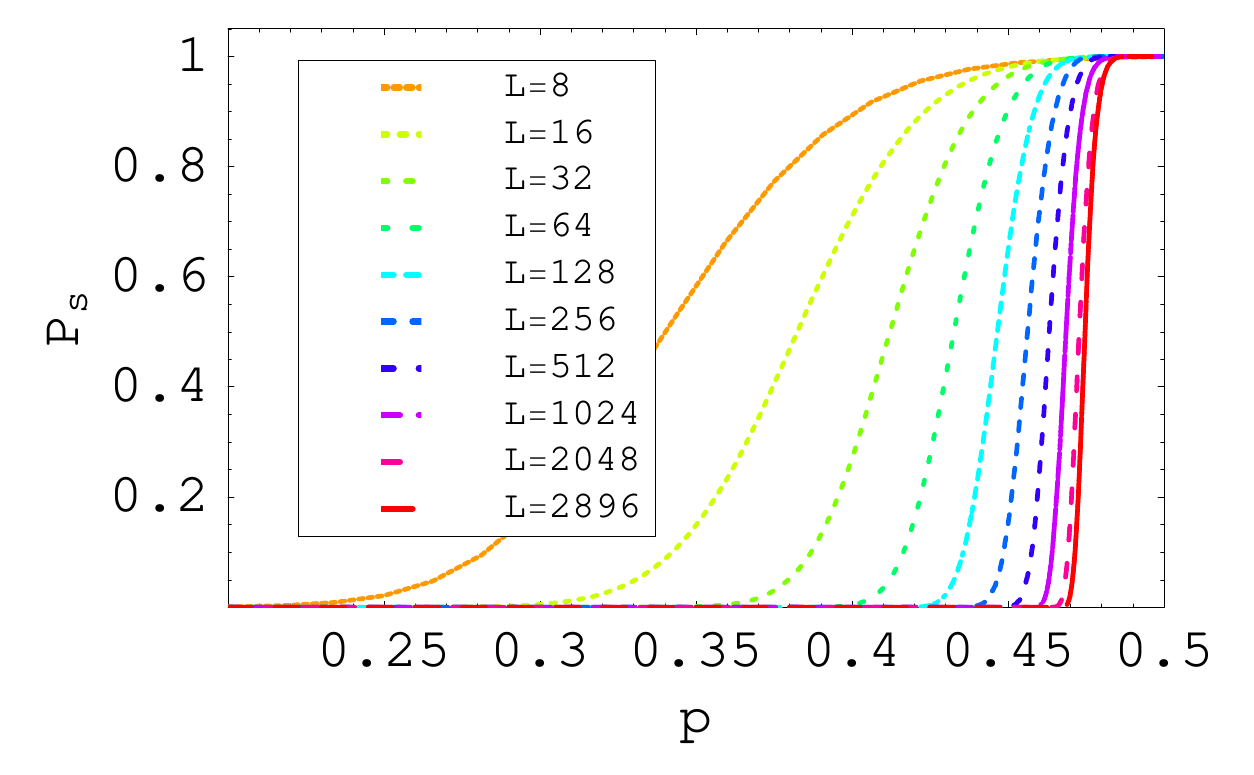}
\caption{The probability of spanning for periodic boundary
conditions for the $16$ NN force-balance model.
Error bars are too small to be seen.}
\label{fig:Z16.periodic.Pspan}
\end{center}
\end{figure}

\begin{figure}[htb]
\begin{center}
\includegraphics[width=0.9\columnwidth]{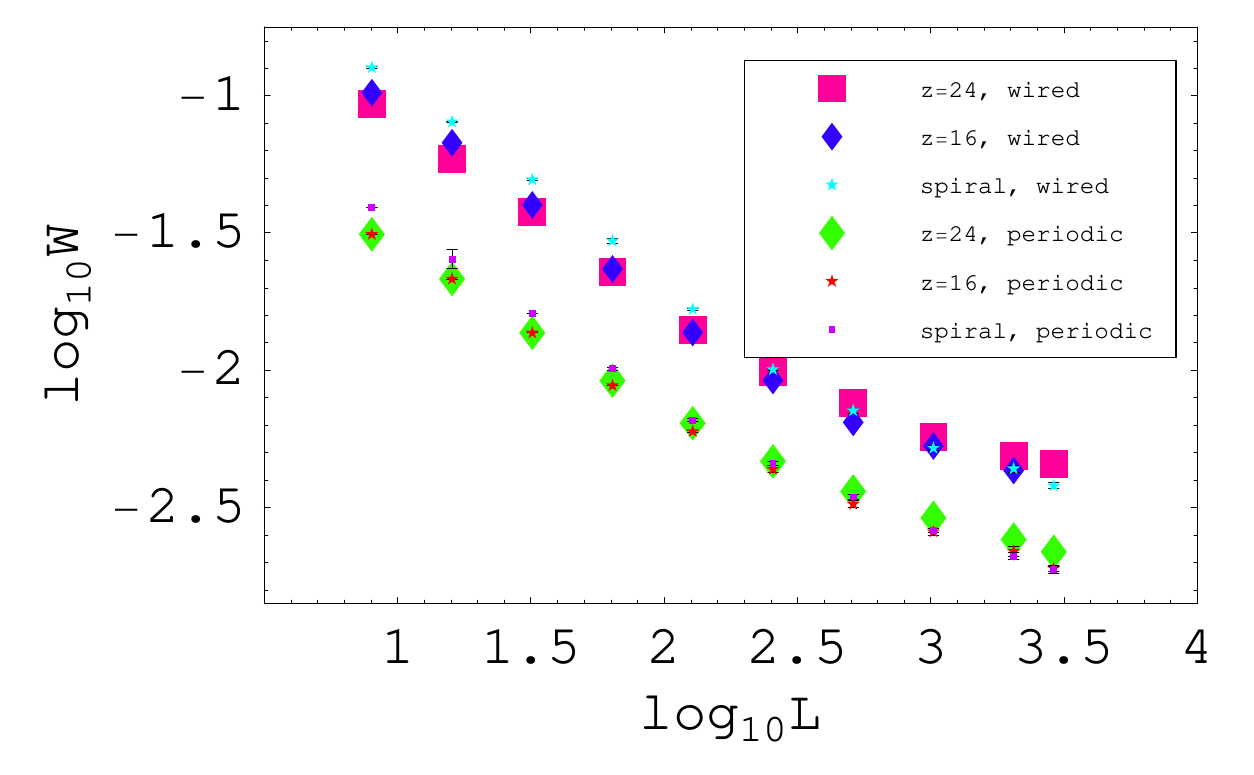} 
\caption{The widths for all three models with both wired and
periodic boundary condition, on a log-log plot. All curves
show clear deviations from power
laws at large system sizes.}
\label{fig:All.Widths.Plot1}
\end{center}
\end{figure}

\begin{figure}[htb]
\begin{center}
\includegraphics[width=0.9\columnwidth]{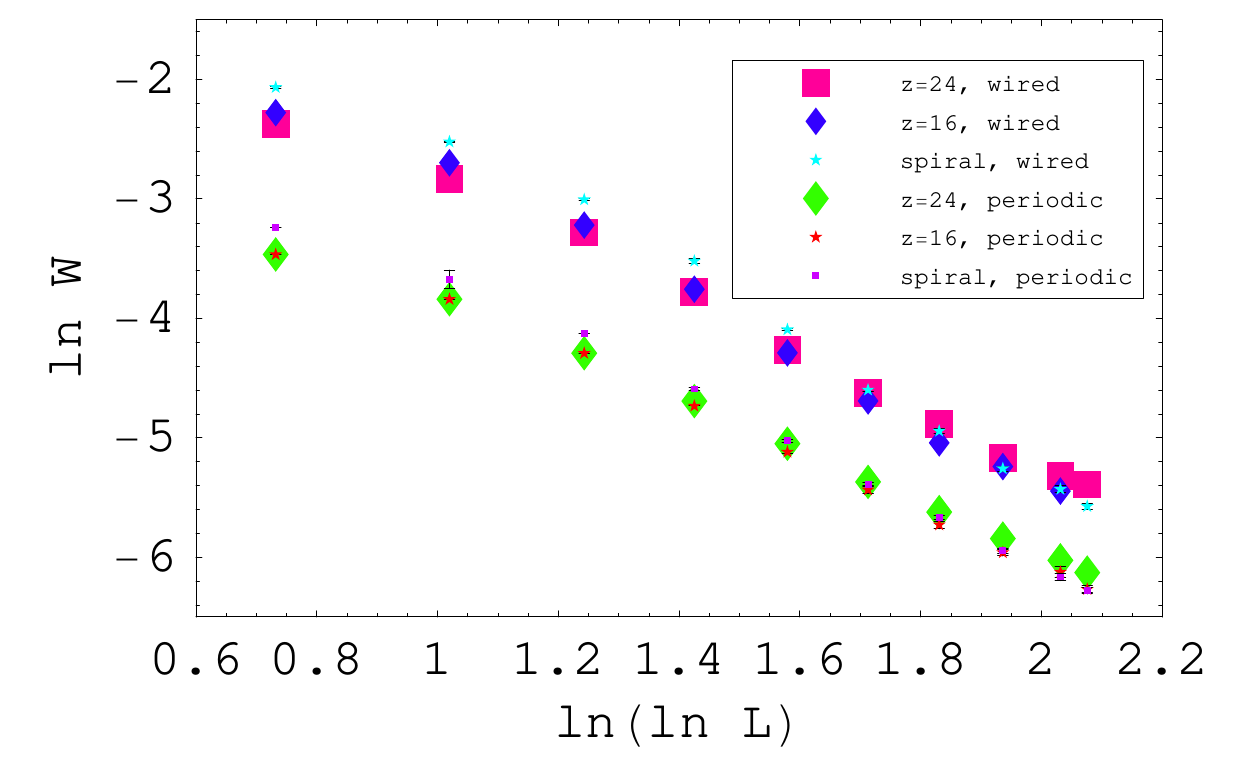} 
\caption{The widths for all three models with both wired and
periodic boundary condition assuming the TBF form.}
\label{fig:All.Widths.Plot2}
\end{center}
\end{figure}

Fig.~\ref{fig:Z16.periodic.Pspan} shows the probability of
spanning for periodic boundary conditions. 
From this data we extract the width of the transition as a
function of $L$, as depicted in Fig.~\ref{fig:All.Widths.Plot1}. We plot both 
periodic and wired boundary conditions on a log-log scale to demonstrate the significant deivations from a power-law growing crossover length. We also plot the same data for the 24NN and spiral models for comparison. As with the 24NN model, the fitting form is clearly not a power law in $L$. 
Figure~\ref{fig:All.Widths.Plot2} shows the widths using the fitting form 
motivated by the TBF result. For the 16NN model, one can extract $\mu=0.35\pm0.01$ for wired boundary conditions and $\mu=0.42\pm0.03$ for periodic boundary conditions.

\begin{figure}[htb]
\begin{center}
\includegraphics[width=0.9\columnwidth]{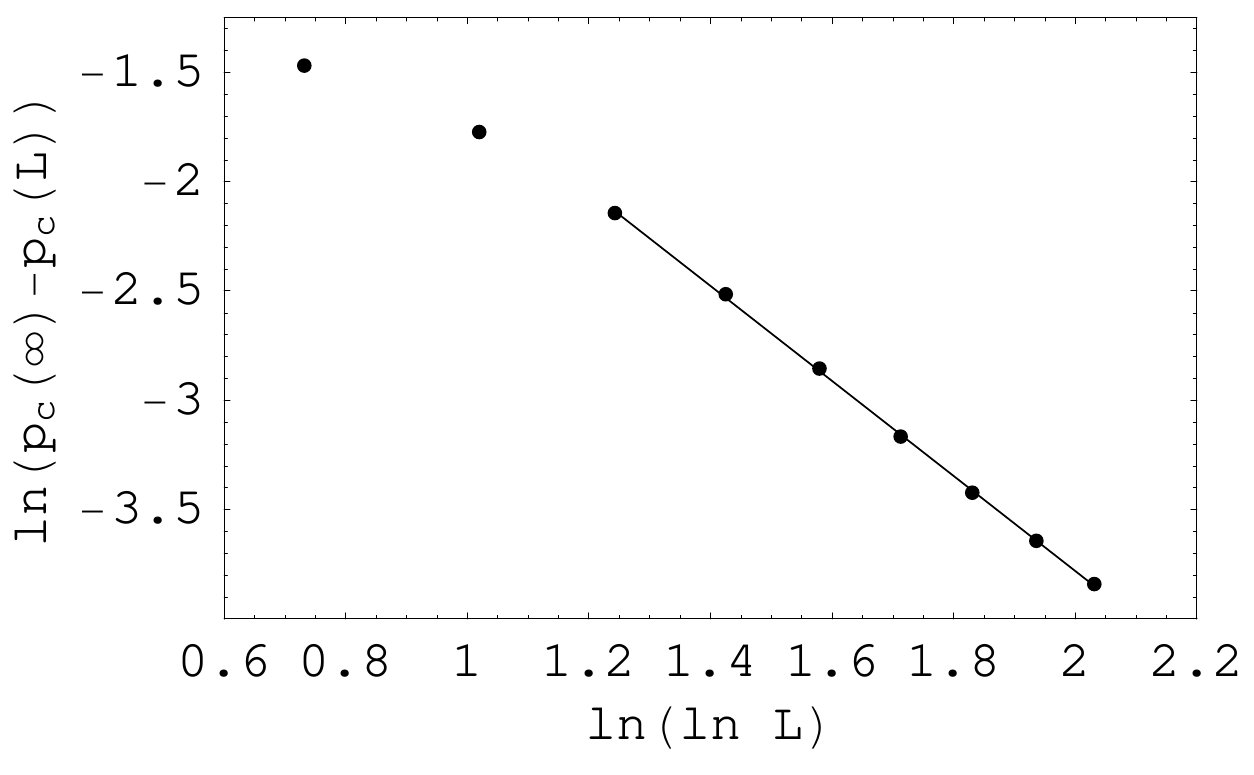}
\caption{Fitting of $p_c(L)$ as a function of $L$ using the
TBF fitting form, for the $16$ NN force-balance model with wired boundary conditions. 
We obtain the best fit with
$p_c(\infty)=0.497\pm0.007$.}
\label{fig:Z16.wired.PCrit.TBFfit}
\end{center}
\end{figure}

In Fig.~\ref{fig:Z16.wired.PCrit.TBFfit}, similarly to
Fig.~\ref{fig:Z24.wired.PCrit.TBFfit}, 
we choose $p_c(\infty)$ so that $p_c(\infty)-p_c(L)$
as a function of $L$ is well fit by the TBF functional form.
We obtain the best fit for $p_c(\infty)=0.497\pm0.007$ for wired boundary conditions, which gives $\mu=0.47\pm 0.07$. For periodic boundary conditions, $p_c(\infty)=0.502\pm0.010$ and $\mu=0.70\pm0.15$. Once again, there is a discrepancy between 
the $\mu$ extracted from the width data and the $\mu$
extracted from the one-parameter fit.

A similar fit (not shown) for the spiral model 
finds $p_c(\infty)=0.690\pm0.008$, and $\mu=0.49\pm 0.08$ for wired boundary 
conditions. 
For the spiral model, Toninelli, {\it et al.}~\cite{TBF.response,Spiral.longer} have proven that 
$p_c(\infty)$ is the same as for direct percolation, 
$p_c(\infty)\cong 0.705$, and found
$\mu=0.64$. 
Our numerical results for $p_c(\infty)$ and $\mu$ are both 
within
two error bars from their exact result,
and a plot of $\ln (p_c(\infty)-p_c(L))$ versus
$\ln(\ln L)$ for the exact result of 
$p_c(\infty)\cong 0.705$
shows noticeable curvature. Also, the $\mu$ extracted from the width data for
wired boundary conditions is $\mu=0.32\pm0.01$ and
$\mu=0.38\pm0.02$ for periodic boundary conditions. 

\begin{figure}[htb]
\begin{center}
\includegraphics[width=0.9\columnwidth]{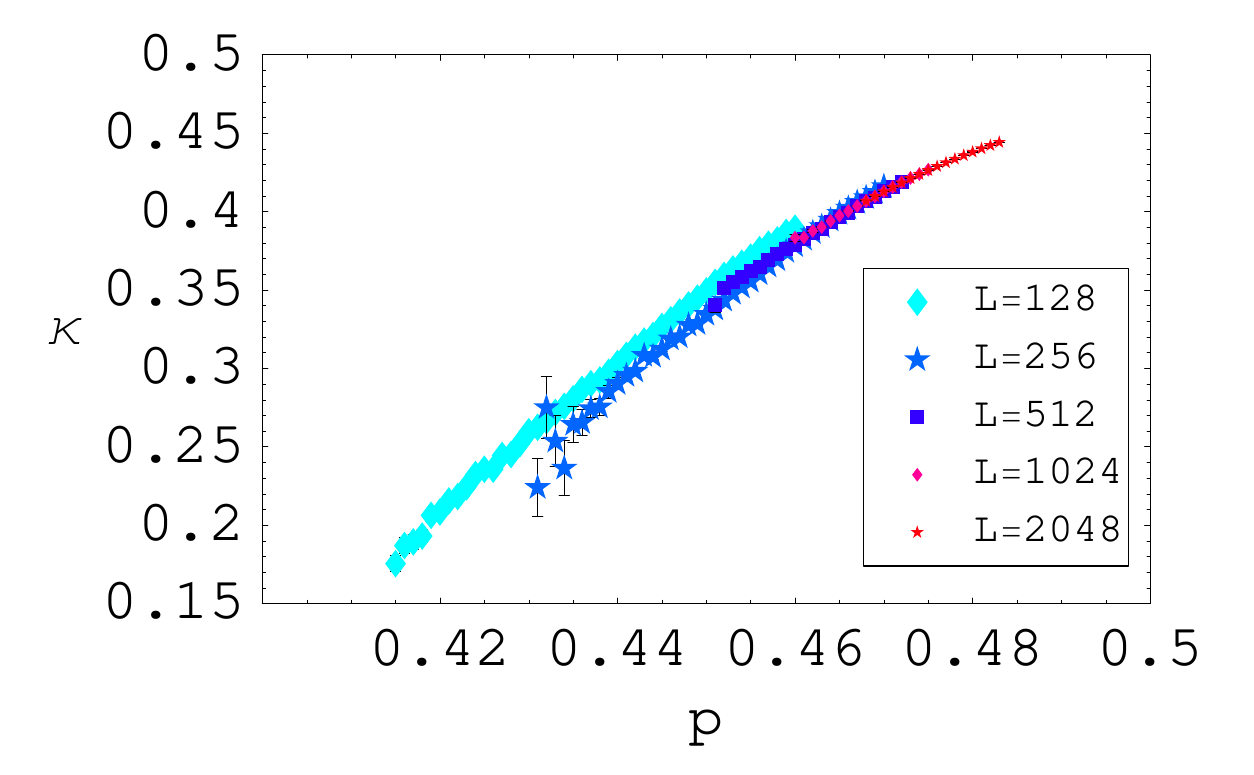}
\caption{Order parameter for the $16$ NN model using wired boundary conditions.}
\label{fig:Z16.wired.OrderParameter}
\end{center}
\end{figure}

Finally, Fig.~\ref{fig:Z16.wired.OrderParameter} shows the 
order parameter $\kappa$ as a function of $p$ for various system sizes for the 16NN model. As in the 24 NN model, the jump in $\kappa$ increases with increasing system size suggesting that the transition is discontinuous. It also appears that $\beta=1$ just above the transition for each individual curve (though there is some overall curvature for the set of curves). 


 

\subsection{Three-dimensional model}

\begin{figure}[htb]
\begin{center}
\includegraphics[width=0.9\columnwidth]{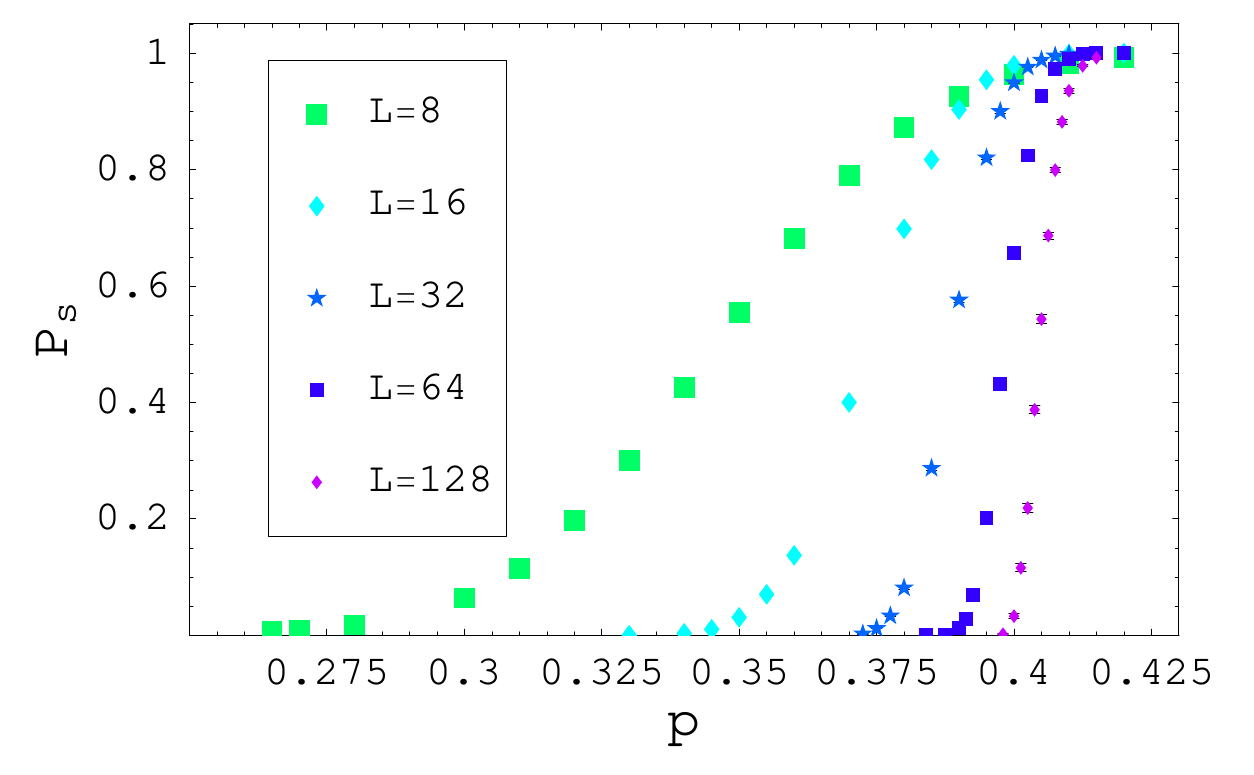}
\caption{Probability of having a spanning cluster in any of three 
directions, as a function of $p$, for different system sizes 
of length $L$, and periodic boundary conditions. This plot 
is for $k=4$ with force-balance on the cubic lattice with 
26 nearest neighbors.} 
\label{fig:Model3D.Pspan}
\end{center}
\end{figure}

Our tests of the 26NN three-dimensional model are not as extensive as in the two-dimensional case, particularly because we have no proof that $p_c<1$ in this 
case and our system size range is limited. 
From the probability of spanning data using periodic boundary conditions, we have extracted $W$ and $p_c(L)$. See Fig.~\ref{fig:Model3D.Pspan}. Deteriming $p_c(\infty)$ with the 
TBF functional form yields $p_c(\infty)=0.433\pm 0.009<1$
(not shown). Moreover, for the one-parameter $p_c(\infty)$
fit, $\mu=0.75\pm0.14$; from the width data,
$\mu=0.37\pm0.01$. In the three-dimensional case the
discrepancy between the two values of $\mu$ for the same
boundary condition is greater than the two-dimensional
force-balance models, where larger system sizes can be explored. For this three-dimensional model, one may expect a double logarthim as opposed a single logarithm with potentially $\mu\cong 0.52$ using the values for three-dimensional directed percolation~\cite{DP}. Given the small range of data, however, it is difficult to discriminate between the two forms. 
 
\begin{figure}[htb]
\begin{center}
\includegraphics[width=0.9\columnwidth]{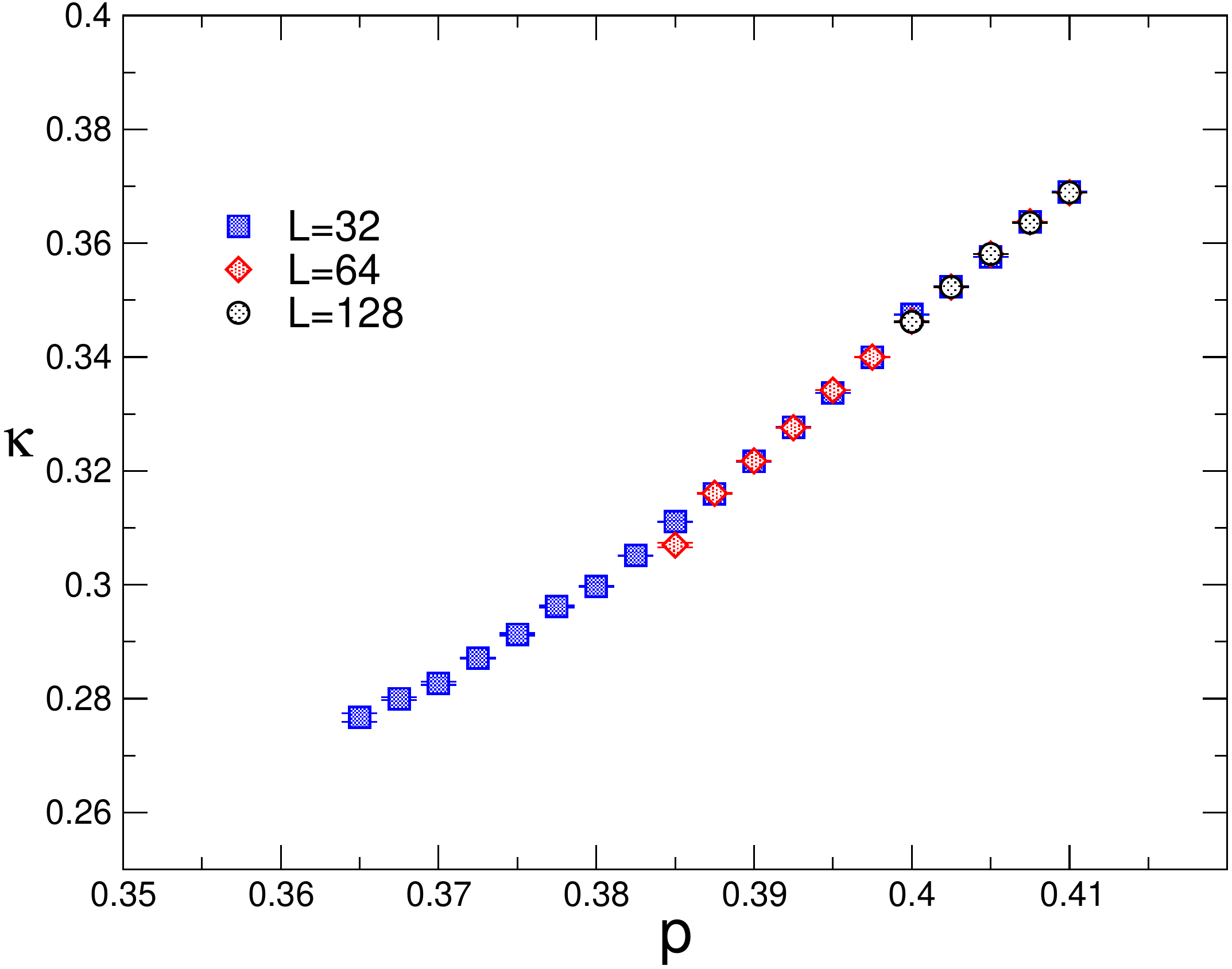}
\caption{Order parameter plot for the three-dimensional model using periodic boundary conditions.}
\label{fig:Model3D.OrderParameter}
\end{center}
\end{figure}

Finally, the three-dimensional transition appears to be
discontinuous in the context of the onset of the infinite
force-balance cluster, as in the two-dimensional cases, since the jump in the order parameter increases with increasing system size. See Fig.~\ref{fig:Model3D.OrderParameter}. Moreover, simulations of the probability of having a force-balance avalanche size $s$ shows the same trends as in Figs. 12 and 13.

\section{Summary and Discussion}
\label{sec:conclusions}

While uncorrelated percolation is very well understood, 
models of correlated percolation are much less so. We have 
presented rigorous and numerical results on several 
force-balance percolation models to help narrow the gap. 
On the rigorous side, we have proven that $p_c<1$ 
for the two-dimensional models. This result places our 
interpretation of the numerical data for the
two-dimensional models on sounder footing. A rigorous argument that 
the force-balance percolation transition is discontinuous, 
at least in two dimensions, is more difficult than in the 
case of jamming percolation. In the jamming percolation 
models the underlying mechanism driving the transition is 
two disjoint, directed percolation processes. These two process scaffold 
upon one another such that the infinite cluster is bulky at the transition. 
The underlying mechanism driving the transition in the force-balance 
models is presumably the same as argued in Sec. IIIB. Models of 
newly-constructed directed percolation processes will be key in constructing 
a rigorous argument for discontinuity in the force-balance case.

Our numerical results for the force-balance avalanches and 
the onset of the infinite force-balance cluster for all 
three force-balance models strongly suggest a discontinuous 
transition. For the force-balance avalanches, the probability 
of having an avalanche size $s$, is broad for intermediate 
avalanche sizes.  There is also a well-defined avalanche 
size at the tail of the distribution that becomes more prominent 
as the system size is increased and as $p$ is decreased 
towards the transition. This suggests a bulky, discontinous,
transition. Looking at 
the usual order parameter
also points towards a discontinuous transition in that the 
jump in the fractional size of the largest force-balance 
cluster at the transition 
increases with increasing system size.  
This trend was reported in Ref.~\cite{SLC} for the 24 NN 
model only.

So now that we know there is a transition (at least for the two-dimensional models) and we surmise that the onset of the infinite force-balance cluster is discontinuous, just as in the case of jamming percolation, is there a quantitative connection with jamming percolation beyond the heuristic arguments provided in Sec. IIIB? 

The results for the mean culling time quantitatively suggest 
that at least the two two-dimensional force-balance models and 
the spiral model are in the same universality class.
We obtained 
mean culling time exponents of $\alpha=1.226\pm0.027$, 
$1.240\pm0.025$, and 
$1.246\pm0.027$, for the 24 NN model, 16 NN model, and 
spiral model, respectively. Since all three exponents are within 
one standard deviation of $5/4$, 
the equivalent exponent in the sandpile model, 
there is potential for a sandpile model-like RG 
treatment, at least for the culling dynamics for both sets 
of models. Further quantitative study of the distribution 
of force-balance and spiral avalanche sizes will provide 
further insights.

The crossover length data also indicates that for the system sizes studied, the mechanisms underlying jamming and force-balance percolation are the same for the two-dimensional models.
The following table summarizes the values obtained for
$p_c(\infty)$ for the three two-dimensional models, and the associated
values of $\mu$:

\begin{eqnarray}
\nonumber
\begin{tabular}{l|l|c|c}
Model & Boundary & $p_c(\infty)$ & $\mu$ \\
\hline
24 NN & wired & $0.414\pm 0.008$ & $0.51\pm 0.09$ \\
16 NN & wired & $0.497\pm 0.007$ & $0.47\pm 0.07$ \\
Spiral & wired & $0.690\pm 0.008$ & $0.49\pm 0.08$ \\
24 NN & periodic & $0.425\pm 0.005$ & $0.76\pm 0.20$ \\
16 NN & periodic & $0.502\pm 0.010$ & $0.70\pm 0.15$ \\
Spiral & periodic & $0.701\pm 0.011$ & $0.81\pm 0.16$
\end{tabular}
\end{eqnarray}

\noindent We see that results for $\mu$ for all three models
with wired boundary conditions are consistent with one
another, although the error bars are large. So, while our
measurements are not as precise, nor as accurate, 
as one would like (the former possibly indicated by the differing values of 
$\mu$ obtained from the width data), the consistency between the three different models is readily apparent. In other words, the models are most likely in the same universality class. We have also
included results for periodic boundary conditions. For
those, the results for $p_c(\infty)$ are within the error
bars of the results for wired boundary conditions;
for periodic boundary conditions, the plots of 
$\ln (p_c(\infty)-p_c(L))$ versus
$\ln(\ln L)$ appear linear for a wider range of
$p_c(\infty)$, resulting in significantly larger error bars.
Finally, the $p_c$'s are independent of the boundary
condition, as expected. 
 
Our force-balance data is indeed {\it inconsistent} with a crossover 
length that diverges as a power law, and is just as consistent 
with the TBF fitting form as  
the spiral model data. 
With the present data, we are 
unable to conclude that the presumably two-dimensional model-independent value of $\mu$ is independent of the boundary 
conditions, though with data for larger systems, and smaller error bars, such a
trend should emerge.
A stronger numerical test of the same underlying mechanism
for jamming percolation and force-balance percolation would
be to look for directed percolation using anisotropic
finite-size scaling, as was done in
Ref.~\cite{TBF.response}. We leave this for future work.

Our correlation length data 
above the critical point for the 24 NN model suggests
a correlation length that grows with system size,
as opposed to a finite one. However, for the largest system sizes, we do 
not generally see the expected
power law correlations at the transition associated with this divergence.  More work is needed to substantiate this trend, which would be inconsistent with a diverging correlation length. Of course, a study of larger systems might reveal a finite correlation length.  
Our current correlation length results are to be contrasted with $4$-core percolation
in four dimensions, where a finite correlation length of
about 10 lattice spacings was found, suggesting a
garden-variety type discontinuous transition driven by
nucleation~\cite{Parisi}. 4-core percolation in four dimensions contains finite
clusters, which provide a backbone for
nucleation. In force-balance percolation, however, there are no
finite clusters so one may expect a more unusual discontinuous
transition. 

Therefore, the scenario for force-balance percolation that is
most consistent with our data is that while the onset of the
infinite force-balance cluster is discontinuous, there is an
exponentially diverging crossover lengthscale, and perhaps a
diverging correlation lengthscale. Our limited data for the
standard correlation length defined in the connected phase
makes it difficult to discern any trend for growth,
exponential or otherwise. In continuous phase transitions,
the crossover length and correlation length diverge in the
same way. With this more unusual transition, it is not
necessarily obvious that the same behaviour should apply. Moreover, we find quantitative agreement with the dynamical exponent for sandpile models not only for the force-balance models but for the spiral model as well, again, suggesting that they all are in the same universality class.  Finally, we expect that three-dimensional versions of jamming percolation and force-balance percolation should exhibit similar behaviour as well, as our data suggests. 

Though the usual order parameter, the fraction of sites in the infinite force-balance cluster, does not appear to be continuous, 
are there other candidates for an atypical continuous order 
parameter? A potential candidate is to look at the 
subset of the force-balance 
spanning cluster where the connectivity is marginal, i.e. 
3-connected. However, we do not find any evidence for a fractal, 
spanning 3-cluster at the force-balance percolation
transition, nor for a fractal, spanning 4-cluster.
Given that $p_c$ is around 0.4, each site has approximately 10 
neigbors at the transition. The $3$-core condition is thus
completely superseded by the vectorial constraint, at
least for this lattice model with many nearest neighbors. 
While the dynamics of culling
suggest a critical sandpile-like model for the removal of redundant 
sites, the removal of the marginal infinite cluster does not. So at this time, 
we have not discovered an order parameter which is 
continuous at the transition.

While the jamming and force-balance percolation models lead 
to a discontinuous transition (provably in the first case, 
and most likely in the latter), the fraction of the sites 
in the infinite cluster grows linearly in $p-p_c$. It would 
certainly be interesting to uncover other finite-dimensional 
models that have 
a discontinuous 
transition in which the fraction of sites in the infinite 
cluster grows nonlinearly in $p-p_c$ above the transition; 
such models would behave more like mean-field models. At this 
point, we know of no such models. It may be that finite-dimensional 
models of correlated connectivity percolation are too simple 
to capture this aspect of jamming, and that one has to define 
forces on the network, as in rigidity percolation. We are 
currently working towards this direction.

MJ and JMS would like to acknowledge very helpful discussions
with Andrea J. Liu and a helpful comment from Cris Moore. 
JMS would like to acknowledge the Aspen Center for 
Phyics where part of this work was completed. Finally, we acknowledge support from NSF-DMR-0645373 and NSF-DMR-0605044.

\end{document}